# Gaia Data Release 3. Summary of the variability processing and analysis

L. Eyer[3], M. Audard[3,5], B. Holl[3,5], L. Rimoldini[5], M. I. Carnerero[15], G. Clementini[6], J. De Ridder[18], E. Distefano[8], D.W. Evans[1], P. Gavras[13], R. Gomel[11], T. Lebzelter[25], G. Marton[16], N. Mowlavi[3], A. Panahi[11], V. Ripepi[26], Ł. Wyrzykowski[24], K. Nienartowicz[4,5], G. Jevardat de Fombelle[3], I. Lecoeur-Taibi[5], L. Rohrbasser[5], M. Riello[1], P. García-Lario[7], A. C. Lanzafame[8,9], T. Mazeh[11], C. M. Raiteri[15], S. Zucker[22], P. Ábrahám[16,17], C. Aerts[18,19,20], J. J. Aguado[10], R. Anderson[21], D. Bashi[22], A. Binnenfeld[22], S. Faigler[11], A. Garofalo[6], L. Karbevska[5,23], Á. Kóspál[16,20,17], K. Kruszyńska[24], M. Kun[16], A. F. Lanza[8], S. Leccia[26], M. Marconi[26], S. Messina[8], R. Molinaro[26], L. Molnár[16,27,17], T. Muraveva[6], I. Musella[26], Z. Nagy[16], I. Pagano[8], L. Palaversa[14,1], E. Plachy[16,27,17], K. A. Rybicki[24], S. Shahaf[12], L. Szabados[16,27], E. Szegedi-Elek[16], M. Trabucchi[30,3], F. Barblan[3], and M. Roelens[3]

(Affiliations can be found after the references)

**ABSTRACT**

*Context.* Gaia has been in operations since 2014, and two full data releases (DR) were delivered so far: DR1 in 2016 and DR2 in 2018. At each data release, more sources, more data types and more analysis results are made available to the public. The third Gaia data release expands from the early data release (EDR3) in 2020, which contained the 5-parameter astrometric solution and the mean photometry for 1.8 billion sources, by providing 34 months of multi-epoch observations that allowed us to probe, characterise and classify systematically celestial variable phenomena.
*Aims.* We present a summary of the variability processing and analysis of the photometric and spectroscopic time series of 1.8 billion sources done for Gaia DR3.
*Methods.* We used statistical and Machine Learning methods to characterise and classify the variable sources. Training sets were built from a global revision of major published variable star catalogues. For a subset of classes, specific detailed studies were conducted to confirm their class membership and to derive parameters that are adapted to the peculiarity of the considered class.
*Results.* In total, 10.5 million objects are identified as variable in Gaia DR3 and have associated time series in $G$, $G_{BP}$, and $G_{RP}$ and, in some cases, radial velocity time series. The DR3 variable sources subdivide into 9.5 million variable stars and 1 million Active Galactic Nuclei/Quasars. In addition, supervised classification identified 2.5 million galaxies thanks to spurious variability induced by the extent of these objects. The variability analysis output in the DR3 archive amounts to 17 tables containing a total of 365 parameters. We publish 35 types and sub-types of variable objects. For 11 variable types, additional specific object parameters are published. An overview of the estimated completeness and contamination of most variability classes is provided.
*Conclusions.* Thanks to Gaia we present the largest whole-sky variability analysis based on coherent photometric, astrometric, and spectroscopic data. Later Gaia data releases will more than double the span of time series and the number of observations, thus allowing for an even richer catalogue in the future.

## 1. Introduction

The Gaia mission (Gaia Collaboration et al. 2016) is contributing to astronomy in many ways. The Gaia astrometry, with its unique precision for so many stars, can be used by photometric surveys to locate stars in the Hertzsprung-Russell Diagram, or to perform kinematic and spatial studies. Furthermore, Gaia with its repeated quasi-simultaneous $G$, $G_{BP}$, and $G_{RP}$ integrated photometric band measurements is also a major player among the multi-epoch photometric surveys, while spectrophotometric RP and BP measurements are a unique feature of Gaia. On top of that, the Radial Velocity Spectrometer (RVS) will provide mean spectra and radial velocity (RV) measurements up to $G_{RVS} \sim 16$ mag, and still unprecedented numbers of epoch spectra and RV values to $G_{RVS} < 14$ mag.

Gaia offers the unique opportunity to study variability of close to 2 billion objects thanks to the $G$, $G_{BP}$, $G_{RP}$ photometric time series, but also to the other spectrophotometric and RVS time series. In previous Gaia data releases, we published 3194 variable stars in DR1 (Eyer et al. 2017) and 550 737 variable stars in DR2 (Holl et al. 2018). In DR3, we enlarge the sample of variable sources by more than one order of magnitude, reaching 10 509 536 variable objects. In this article, we elaborate on the latter sample, and group the sources into 35 variability classes (variable types and subtypes): 34 different stellar variability (sub)types plus an additional group of variable sources identified as extragalactic, namely active galactic nuclei (AGN)/quasars.

The Gaia variability processing has many ramifications that are disseminated in several other articles. The variety of results makes the situation somewhat intricate. We give here a summary of the results (see Table 1) and we provide associated queries in Appendix B. In total, 14 198 022 sources (variable sources, galaxies, sources in the Gaia Andromeda Photometric Survey,

**Table 1.** Summary of Time-Domain information in DR3.

| Counts | Description |
|---:|---|
| 14 198 022 | All processed (including spurious variability from galaxies) and GAPS |
| 12 960 900 | All variables and galaxies |
| 12 428 245 | All classifications: classification of variables and galaxies |
| 11 754 237 | `vari_summary`: all variables + GAPS (with overlaps), i.e. number of sources having time series |
| 10 509 536 | All variable sources |
| 9 976 881 | Classifications of variables from the supervised classifier |
| 2 451 364 | Classification of galaxies (spurious variability, in the galaxy candidates table) |
| 5 834 543 | All variables with specific studies |
| 1 257 319 | GAPS, among which 12 618 published variable sources and 7579 galaxies |
| 1898 | Variable stars (RR Lyrae stars and Cepheids) with radial velocity time series |

GAPS) have DR3 outputs processed by the Gaia Data Processing Centre of Geneva (DPCG).

In the context of 'real-time' variability analysis of the Gaia photometric data it is important to mention that the Gaia Science Alert system (Hodgkin et al. 2021) has published 10 765 alerts between June 2016 and December 2019. A subset of 2 612 sources detected between 25 July 2014 and 28 May 2017 are published along with Gaia DR3.

Along with the time series for variable objects, DR3 also includes time series and statistical information for all 1 257 319 sources observed within a cone of 5.5° opening directed towards the Andromeda galaxy (GAPS, Evans et al. 2022), independent of their variability status. In the GAPS data set, 12 618 sources are flagged as variables.

Furthermore, the signal of the processed sources is known to be affected by scan angle orientation effects inducing spurious variability (Holl et al. 2022). Such effects allowed us to identify 2 451 364 galaxies using Machine Learning supervised classification techniques, which are included in DR3 in a dedicated catalogue.

The structure of the article is as follows: In Sect. 2, we present the properties of the data from which DR3 results are derived. In Sect. 3, we describe the methods used and the obtained results. In Sect. 4, we present the merged view of the variability types. We also show properties of the variability types from the classification and specific study perspectives. Furthermore we add a comment on differences that can be found in different archive tables. Section 5 provides completeness and contamination estimates for the different variability classes. Section 6 presents the colour-magnitude diagrams for variable stars in the Large and Small Magellanic clouds. In Sect. 7, we present an ongoing collaboration between Gaia and TESS missions related to detection of exoplanetary transits. In Sect. 8, we finish with some concluding remarks.

## 2. Properties of Input Data

### 2.1. Photometric time series

Photometric data in the $G$, $G_{BP}$ and $G_{RP}$ photometric bands were used from July 25, 2014 to May 28, 2017, i.e. 34 months. The sampling cadence depends on the spacecraft's "scanning law", that is, the mode by which Gaia spins and precesses around its rotation axis. During the 34 months of observations processed here, two different scanning laws were used: the Ecliptic Pole Scanning Law (EPSL) during the first 28 days of the operations and then the Nominal Scanning Law (NSL), see section 1.3.2 of the Gaia DR3 documentation (van Leeuwen et al. 2022). The implementation of these two scanning laws resulted in higher cadences and numbers of observations for objects near the ecliptic poles and the ecliptic latitude $\beta = \pm 45°$, compared to elsewhere on the sky.

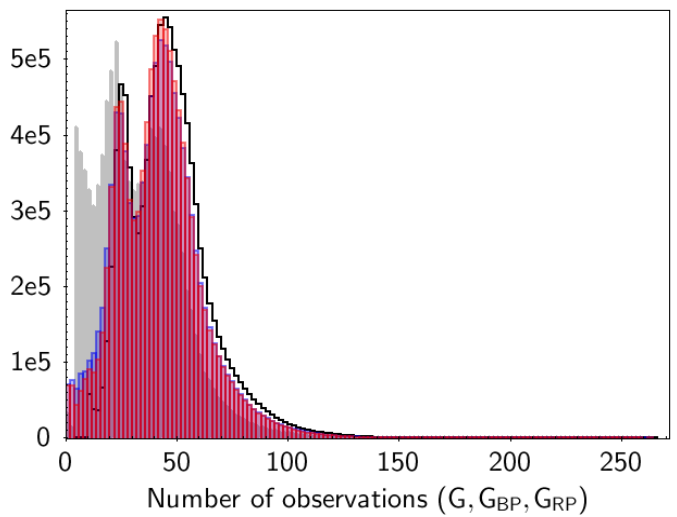

**Fig. 1.** Histogram of DR3 photometric FoV observations for the variables in the $G$ (black), $G_{BP}$ (blue) and $G_{RP}$ (red) bands. The median numbers of measurements of G, BP and RP are 44, 40 and 41 though it extends up to 265 in the G band. In grey, we show the histogram of measurement numbers for the $G$ magnitude of a random sample of 10.5 million stars. We see that the variability analysis favours a high number of measurements, as expected.

We analysed 1 840 947 142 sources, corresponding to the number of input sources with 5 or more "valid" (i.e., non-NaN) measurements in the $G$ band. The Gold, Silver and Bronze photometric sets (see Riello et al. 2021) were used. The CCD data were only used for the short timescale analysis, although they were not made public in Gaia DR3. The total number of photometric measurements in $G_{BP}$, $G$ (Field-of-View and Per-CCD) and $G_{RP}$ photometric bands amounts to 367 billion.

Eventually, the total number of sources published in DR3 is 1 811 709 771, with 11 754 237 sources having epoch photometry time series and 1898 having RV time series. We repeat here that the variable source data set comprises 10 509 536 entries.

In Fig. 1, we show the number of Field-of-View (FoV) measurements per band ($G$, $G_{BP}$, $G_{RP}$) for the published variable sources, with, in grey, a random sample for the $G$ magnitude of the same amount of Gaia DR3 variable sources. The distribution of variable sources has two marked peaks: one at 25, which corresponds to the many sources in the Galactic bulge, while the other at 45 corresponds to the typical number from the scanning law. Figure 2 presents the histogram of mean $G$ magnitudes for the variable sources compared to a random sample (grey). As we can expect, the variability analysis is more efficient for the bright stars. In general, the fraction of variable stars is about 0.6% for the entire range of magnitudes, but it is higher than 5% up to magnitude 14.

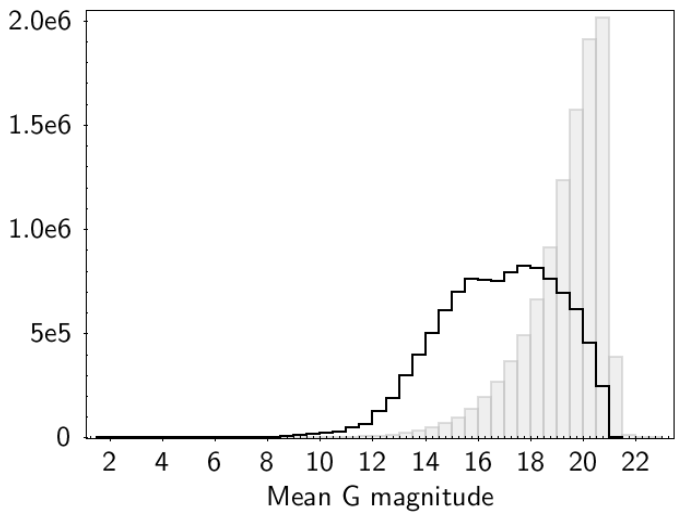

**Fig. 2.** Histogram of the mean magnitudes of the 10.5 million variable sources (black line). For comparison the histogram (grey) for a random selection of 10.5 million sources among the 1.8 billion.

### 2.2. Radial velocity time series

For a subset of RR Lyrae and Cepheid stars, time series of RVS radial velocities were analysed in addition to the photometric data. The output parameters of the radial velocity analysis can be found in table `gaia_dr3.vari_cepheid` and `gaia_dr3.vari_rrlyrae`, when available. In addition, the statistical parameters of these time series (e.g., mean, median, etc) can be found in `gaia_dr3.vari_radvel_statistics` and the related radial velocity time series in `gaia_dr3.vari_epoch_radial_velocity` for 1898 sources, totalling 43 298 measurements. We refer to Clementini et al. (2022) and Ripepi et al. (2022) for more details about the selection and RV analysis.

### 2.3. Spectrophotometric time series

The work package of long-period variables used time series of the red band of the spectrophotometer. The pseudo-wavelength spectra were cut at the edges to exclude noise. The RP spectra allowed us to determine whether these variables were C-rich or O-rich. We refer to Lebzelter et al. (2022) for more details.

## 3. Variability Processing and Analysis steps and overview of the DR3 results

We provide below the main processing and analysis steps, which are shown in Fig. 3. More detailed information can be found in the Gaia documentation and in the articles dedicated to specific variability types. Furthermore, we give an overview of the results for the variable sources published in DR3.

### 3.1. Data Cleaning

Despite the high quality of the Gaia products, the raw time series (photometric or radial velocities) need some cleaning for proper variability analysis. A set of operators applying different cleaning steps are applied. For a detailed description, we refer to Holl et al. (2018) and the DR3 documentation (Rimoldini et al. 2022b).

In DR3, the chain of operators for photometric analysis was modified compared to DR2 (see DR3 documentation, section 10.2.3): a new operator called MultiBandOutlierRemovalOperator (MORO) was introduced. MORO takes advantage of the quasi-simultaneous observations in the $G$, $G_{BP}$, and $G_{RP}$, and thus, the 3-band transits have the same transitid qualifier.

`MORO` was inserted after `GaiaFluxToMagnitudeOperator`, which converted raw fluxes into magnitudes using the band zero points (Riello et al. 2021).

The new operator detects *candidate* outliers in the 3 bands for each transit, based on the ratio iqrMedianMag defined as

$$\text{iqrMedianMag} = \frac{|\text{mag} - \text{median}(\text{mag})|}{\text{iqr}(\text{mag})} \quad (1)$$

where mag is the transit magnitude in one band, median(mag) is the median value, and iqr(mag) is the interquartile range of the time series. As initial step, the *candidate* outliers in each band separately are flagged if iqrMedianMag > iqrMedianMag$_0$, where iqrMedianMag$_0$ is a band-specific threshold. For simplicity and after analysis, the same threshold of iqrMedianMag$_0$ = 3 was used for $G$, $G_{BP}$, and $G_{RP}$. The direction (i.e., brighter or fainter than the median) for each band is determined and used to decide whether the band outlier candidate is a real outlier or a non-outlier (e.g., if $G$ and $G_{BP}$ and/or $G_{RP}$ are in the same directions, they are not considered as outliers; while outlier candidates in both $G_{BP}$ and $G_{RP}$ but not $G$ are considered as real outliers). If only one band has a candidate outlier and measurements exist (in the same transit) in the other bands, the threshold for candidate outlier is reduced to one and the same procedure is followed as described above.

`MORO` also includes a detection of large outliers using ratios of iqrMedianMag in $G_{BP}$ and/or $G_{RP}$ divided by the iqrMedianMag in $G$. If the ratio is above five, the outlier in $G_{BP}$ and/or $G_{RP}$ is considered as a real outlier. In addition, `MORO` has a step looking into the deviations of magnitude errors, however applied independently in each band, using:

$$\text{iqrMedianMagError} = \frac{|\text{magError} - \text{median}(\text{magError})|}{\text{iqr}(\text{magError})} \quad (2)$$

and a threshold of 20 in $G$ and 10 in $G_{BP}$ and $G_{RP}$. For a detailed description of the operator, we refer to Section 10.2.3 in the DR3 documentation.

After `MORO`, the operator `TimeIntervalFilter` (TIF) was applied to cut specific OBMT revolution periods when the data were of poor quality, based on time series for a selection homogeneous sample of sources across the sky. As diagnostic, we used `iqrMedianMag` to detect systematic deviations above the empirical threshold of 3. The intervals often referred to calibration issues, decontamination periods, etc. Intervals were introduced for each band separately.

In DR3, the final operator used for determining statistical parameters and for the GVD and classification steps was `ExtremeErrorCleaningMagnitudeDependentOperator` (in short EECMDO). This is different from DR2 where the operator `RemoveOutliersFaintAndBrightOperator` (ROFABO) was the last step. However, some Specific Object Studies (SOS) packages used ROFABO to clean further the individual time series for their analysis (e.g., the work packages for Cepheids and RR Lyrae stars, for exoplanetary transits, for eclipsing binaries, using package-specific parameters for the operator).

We note that the variability flag, `rejected_by_variability`, provided in the archival data links for epoch photometry refers to the EECMDO step; therefore, end users cannot determine from archival data the time series used by work packages that used ROFABO for their analysis (in a similar fashion as was done in DR2). We also emphasize that the flag rejected_by_photometry refers only to the mean photometry (as available in gaia_source) and was not used by the variability analysis.

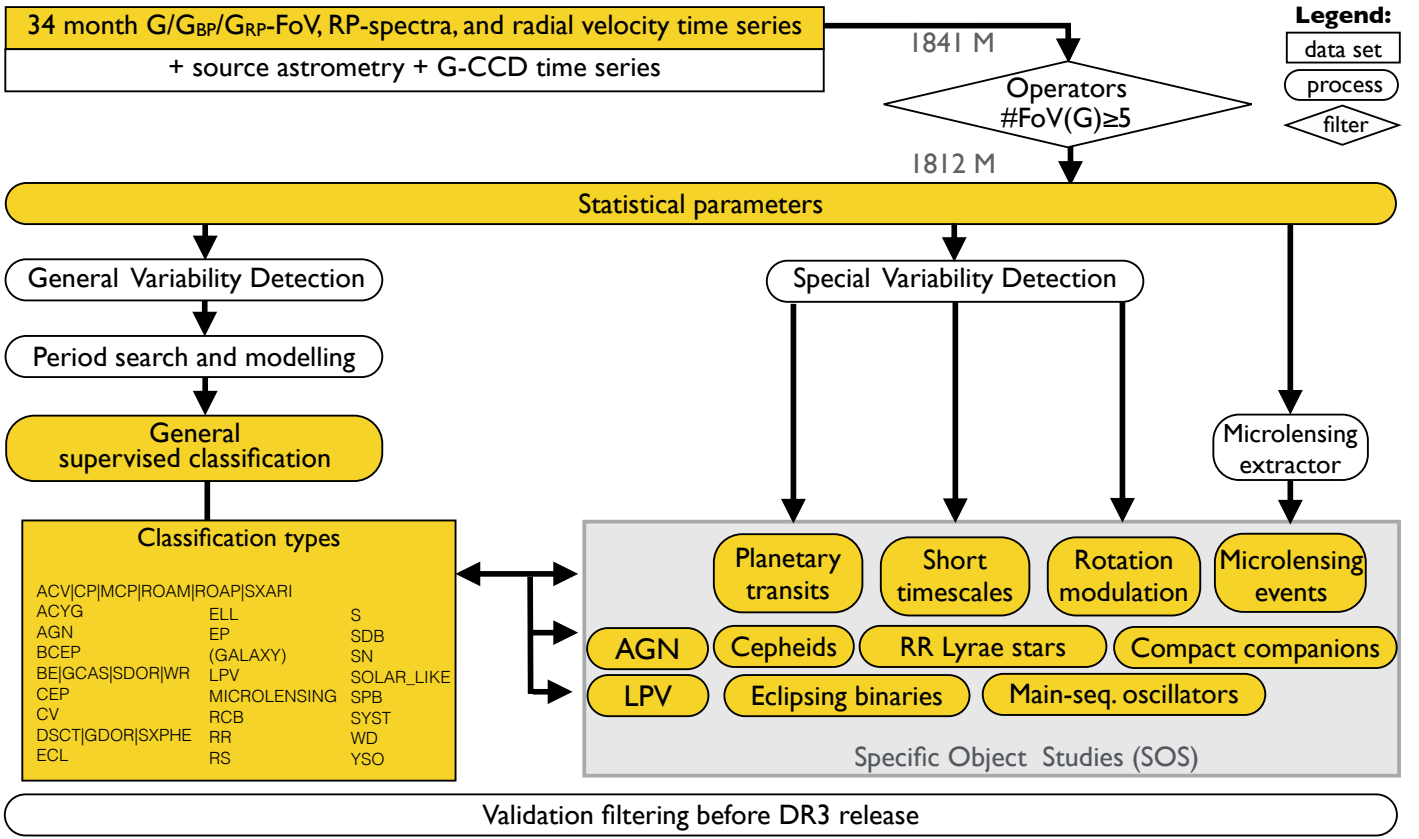

**Fig. 3.** Overview of the DR3 variability analysis pipeline. The steps are data cleaning (operators) and statistical parameters calculations. Then the pipeline splits into two parts: The General Variability Detection (GVD) and the Special Variability Detection (SVD). A supervised classification is applied with 24 variability types. Tailored analyses are done for 11 specific variability types, called Specific Object Studies (SOS) packages.

The thresholds used by the EECMDO operator (see Holl et al. 2018 for details) were computed based on DR3 photometry, using the Gold, Silver, and Bronze sets. Of particular mention, while only upper thresholds were used for $G_{BP}$ and $G_{RP}$ in DR2, we used additional lower thresholds in DR3. The thresholds can be found in Section 10.2.3 of the DR3 documentation (Rimoldini et al. 2022b). The ExtremeValueCleaning operator cut transit magnitudes above 24.5, 24.0, and 23.5, in $G$, $G_{BP}$, and $G_{RP}$, respectively.

The cleaned time series were then input in the statistics package to calculate statistical parameters, provided in the DR3 table gaiadr3.vari_summary, which also contains the information of the presence of sources in other variability tables (e.g., boolean column in_vari_cepheid means that a source can be found in gaiadr3.vari_cepheid). Time series are available for 10 509 536 variable objects (i.e., they appear in one of the variability tables as indicated by gaiadr3.vari_summary, but also in gaiadr3.gaia_source.phot_variable_flag indicated as 'VARIABLE') and for objects in the Gaia Andromeda Photometric Survey (gaiadr3.gaia_source.in_andromeda_survey, see Evans et al. 2022). We emphasize that 1 244 701 sources in GAPS that were not published as variable in DR3 also have entries in gaiadr3.vari_summary. The epoch photometry is available via the DataLink in the Gaia archive, together with reject_by_variability boolean flag, which indicates whether the variability pipeline did not use the transit in the EECMDO step of the processing.[1] Users can also determine the availability of epoch photometry with gaiadr3.gaia_source.has_epoch_photometry.

We note that, in the archive, some sources have statistical parameters in gaiadr3.vari_summary with empty entries in $G_{BP}$ or $G_{RP}$ bands[2], due to the fact that our operators eventually removed all points in a band upstream from EECMDO. This is the case for 20 788 sources in $G_{BP}$ and 19 564 for $G_{RP}$. There are 19 410 cases where there are both no $G_{BP}$ and $G_{RP}$ statistical parameters, while there are 1378 and 154 with null entries in $G_{BP}$-only and $G_{RP}$-only, respectively.

The RV time series were also cleaned for outliers, using an adapted version of the ROFABO operator. Instead of using the interquartile range, the median absolute deviation was used:

$$\text{madMedianRadVel} = \frac{|\text{radVel} - \text{median}(\text{radVel})|}{\text{mad}(\text{radVel})} \quad (3)$$

where radVel is the transit radial velocity. Outliers were flagged when they exceeded a threshold of madMedianRadVel > 300.

---
[1] reject_by_photometry is a different boolean flag, which reflects whether the transit was used to calculate the mean flux and magnitude provided in gaiadr3.gaia_source (i.e., not the mean magnitude in gaiadr3.vari_summary)

[2] Overall, all these cases can be found, e.g., using the ADQL command select * from gaiadr3.vari_summary where num_selected_bp IS NULL or num_selected_rp IS NULL

Users can alternatively determine the availability of RV time series with `gaiadr3.gaia_source.has_epoch_rv`.

As mentioned above, the RP time series spectra used by the long-period variable work package were cut at the edges (7 array indices on both sides). No additional cleaning was performed. Note that these spectra are *not* published in DR3.

### 3.2. Variability detection

During the successive data releases, we have used several approaches to the variability detection. The variability pipeline reflects the different approaches through General Variability Detection (GVD) and Special Variability Detection (SVD) paths (see Fig. 3).

For the GVD path, the most classical approach relies on hypothesis testing using the uncertainty of individual FoV measurements. As explained in Eyer et al. (2017), this was difficult to pursue since a re-scaling of the uncertainty and the presence of outliers have to be taken into account. For Gaia DR1, we used a modification of this approach to establish empirical p-value determinations from different statistics of constant stars. As the variable star analysis was focused around the South Ecliptic pole, constant stars were selected only from the OGLE surveys.

For Gaia DR2, a binary classifier (Random Forest) was trained with about 29 000 known objects from the literature to identify variable and constant sources across the whole sky (see sect. 7.2.3 of Eyer et al. 2018). Nevertheless, the characterisation of the constant class was incomplete and required a semi-supervised step to fill a lack of representatives in the magnitude distribution.

For Gaia DR3, considering the improved photometry in each data release, we chose not to rely on the constancy determined by other surveys and considered as 'constant' (i.e., generally not considered for DR3 variability analysis) the 75% least variable objects, selected as a function of magnitude in bins of 0.1 mag (see sect. 10.2.3 of Rimoldini et al. 2022b). Only a few of the trained variability types occasionally extended below this threshold. A classifier was trained for GVD, with about 60 000 sources representing all of the targeted variability types and a similar amount of least variable objects as defined above, so that exceptions could be made for specific low-amplitude classes (Gavras et al. 2022). The true fraction of variable sources from Gaia is expected to be lower than 25%, so this approach is rather inclusive and permissive. The distinction between constant and variable is not published in the ESA archive.

The SVD path (cf. Eyer et al. 2017) allows us to detect variability types that are irregular or require a special preliminary treatment to identify initial candidate objects, which are then analysed in more detail in the related SOS package. Three packages use this route: planetary transits, short timescales, and rotation modulation.

We note that an additional path was used for microlensing, where a specific, dedicated extractor identified candidates as input for a focused treatment in the SOS package.

### 3.3. Classification

Supervised classification was performed on the variables detected in the GVD path after a period search and modelling of the time series. This package provided ranking scores for the GVD variables for a total of 24 class groups (in addition to galaxies): $\alpha^2$ Canum Venaticorum or (Magnetic) Chemical Peculiar or Rapidly Oscillating Am/Ap-type[3] or SX Arietis star (ACV|CP|MCP|ROAM|ROAP|SXARI), $\alpha$ Cygni-type star (ACYG), Active Galactic Nucleus (AGN), $\beta$ Cephei variable (BCEP), B-type emission line star or $\gamma$ Cassiopeiae or S Doradus or Wolf-Rayet star (BE|GCAS|SDOR|WR), Cepheid (CEP), Cataclysmic Variable (CV), $\delta$ Scuti or $\gamma$ Doradus or SX Phoenicis star (DSCT|GDOR|SXPHE), Eclipsing binary (ECL), Ellipsoidal variable (ELL), Exoplanet (EP), Long-period variable (LPV), Microlensing event (MICROLENSING), R Coronae Borealis variable (RCB), RR Lyrae star (RR), RS Canum Venaticorum variable (RS), Short timescale object (S), Sub-dwarf B star (SDB) of type V1093 Her or V361 Hya, Supernova (SN), Solar-like star (SOLAR_LIKE), Slowly Pulsating B-star variable (SPB), Symbiotic Star (SYST), pulsating White Dwarf (WD), and Young Stellar Object (YSO). For a detailed description of classification and its output, please refer to Rimoldini et al. (2022a). Here, we provide a summary of the high-level variability types in a piechart, cf. Fig. 4.

Data sets from the literature were assembled in order to train the classifiers, Gavras et al. (2022) provides a detailed explanation of the cross-match with Gaia of different catalogues from the literature. Rimoldini et al. (2022a) cross-matched the sources found in a selection of 152 catalogues with Gaia data. This cross-matched literature catalogue with Gaia gathers 7.8 million sources among which 4.9 million variable sources representing over 100 variability (sub)types, 1.2 million non-variable objects, 1.7 million galaxies.

The DR3 results of classification for the variable objects, i.e., 9 976 881 sources, are available in the `vari_classifier_result` table, together with tables `vari_classifier_name` and `vari_classifier_class_name`, which describe the classifier and the class names, in a similar fashion as in DR2.

Classification of galaxies   Shortly after the Gaia second data release, it was realised that galaxies surprisingly contaminated the RR Lyrae star sample. To mitigate this problem, a training set of galaxies was introduced in the supervised classification to lower the contamination of the RR Lyrae candidates. The origin of the spurious signal was understood and can be found in Holl et al. (2022). However, this initial problem can turn into a method to detect galaxies. The supervised classification produced a list of 2 451 364 galaxy candidates, ironically the largest output of the variability classification. As this signal is spurious, there are, in general, no time series associated with these objects. However, there are 7579 light curves published in the pencil beam (Evans et al. 2022). Furthermore, other variable sources could be galaxies with a spurious variable signal (classified as in another class, e.g. in AGN). The galaxies are published as part of a separate table, `galaxy_candidates`.

The class YSO detected 79 375 young stellar objects thanks to their variability. Cross-matches with several catalogues were used to assess completeness and contamination. Most of the YSOs found are located in the direction of known star forming regions and in the Galactic mid-plane. Despite a completeness at the level of a percent, about 40 0000 new YSO candidates were found. More details can be found in (Marton et al. 2022).

---

[3] the roAp, roAm (rapidly oscillating stars) were grouped by the classifier, because most probably the Ap star variability is dominating the small oscillations.

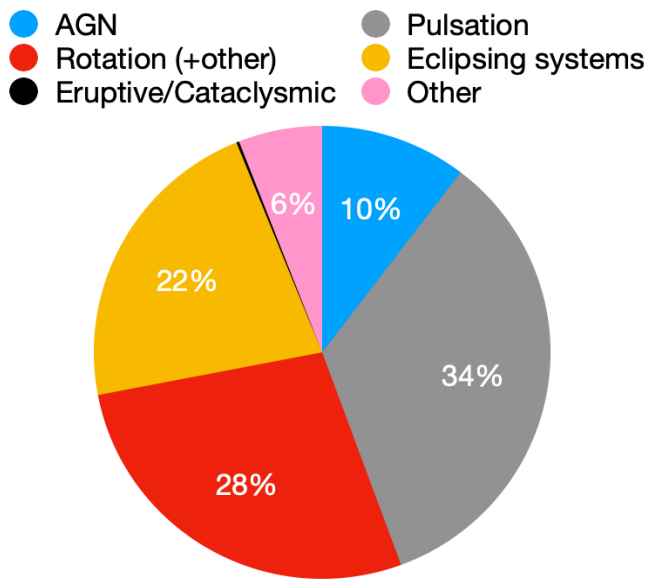

**Fig. 4.** Pie chart of the main causes of variability from the classification output. The groups are formed by the following types: AGN, Rotation (ACV/.../SXARI, ELL, RS, SOLAR_LIKE), Eruptive/Cataclysmic (BE/.../WR, CV, RCB, SN), Pulsation (ACYG, BCEP, CEP, DSCT/GDOR/SXPHE, LPV, RR, SDB, SPB, WD), Eclipsing systems (ECL, EP), Other (MICROLENSING, S, SYST, YSO).

## 3.4. Specific Objects Studies

The DR3 release offers a significant expansion on types of variability compared to DR2, increasing the number of SOS tables from 5 to 11. Below we summarise the different SOS tables.

### 3.4.1. Active Galactic Nuclei

Being magnitude-limited to about $G$ = 20.7 mag, Gaia is not only observing stars but also extragalactic sources such as AGN. AGN are known to be variable, and these can be selected and identified by the supervised DR3 general classification (Rimoldini et al. 2022a). For the AGN specific studies (for details Carnerero et al. 2022), there are two origins in the Gaia DR3 `vari_agn` table: (1) a cleaned list from the classifiers and subsequent filters, and (2) peculiar sources that did not pass filtering, but were considered of high interest, including known blazars and lensed AGN.

The goal of this work was to select a high-purity sample utilising variability properties, including the structure function (Simonetti et al. 1985) and the variability metrics by Butler & Bloom (2011), and further properties, like colours, astrometric parameters, and environment. In addition, as a demonstration, Carnerero et al. (2022) estimated the time lag between the photometric time series of the multiple images of a lensed AGN. The Gaia sampling is often sparse; however, Gaia time series could be systematically used from surveys of lensed quasars to complement their data.

The table `vari_agn` presents 872 228 AGN candidates with more than 21 000 new identifications.

### 3.4.2. Cepheids and RR Lyrae stars

The Cepheids and RR Lyrae stars were present already in the first two Gaia Data Releases. At each release, updates to the pipeline and its output were done for RR Lyrae stars (`vari_rr_lyrae`, see Clementini et al. 2022), and Cepheids (`vari_cepheid`, see Ripepi et al. 2022). From the Fourier decomposition, astrophysical parameters of importance such as the metallicity could be derived.

The RR Lyrae star catalogue nearly doubles the one from Gaia DR2. It is the largest catalogue over the whole sky with 270 905 entries, reaching the very faint limit of Gaia with the mean $G$ mag of 21.14.

The Cepheid sample is reaching 15 006[4] with 5221, 4663, 4616, 321 and 185 in the Large Magellanic Cloud (LMC), the Small Magellanic Cloud (SMC), M31, M33, and field stars/small Milky Way satellites. From the literature, several reclassifications were done into the Cepheid class, amounting to 327 objects, and 474 Cepheid candidates were not reported in the literature.

In addition, RV time series for 1898 sources are included.

To demonstrate the beautiful Gaia data set with $G$, $G_{BP}$, $G_{RP}$ and RV measurements, an Image of the Week was published presenting the Hertzsprung progression. [5]

### 3.4.3. Compact companions

Gaia is an efficient detector for binaries with three observables (astrometry, photometry, spectroscopy). The variability pipeline also aimed to detect binaries with a compact companion, i.e., a black hole, a neutron star or a white dwarf. The approach used was to select ellipsoidal variability which could be caused by a compact object. Gomel et al. (2021) showed that the identification could be achieved by a method based on modified robust minimum mass ratio (mMMR) that can be derived directly from the ellipsoidal amplitude without knowledge of the primary radius or mass. The nature of the compact companion depends then on the estimated minimum mass of the secondary. The approach is conservative since the mMMR is always much smaller than the actual mass ratio in the binary, therefore candidates with large mMMR have a high probability to contain a compact companion.

Candidates for binary ellipsoidals with compact companions are studied in `vari_compact_companion`, see Gomel et al. (2022). Out of more than 6000 candidates, 262 have mMMR larger than unity, with larger probability of bearing a compact companion. These should be studied with radial velocities for necessary confirmation.

### 3.4.4. Eclipsing binaries

Another addition in DR3 is the SOS output for eclipsing binaries in `vari_eclipsing_binary`, see Mowlavi et al. (2022). This is the first Gaia catalogue of eclipsing binaries, containing over two million candidates. The geometry of the $G$-band light curves is modelled with a combination of Gaussian functions to approximate the eclipses and a sine function to model any ellipsoidal variability due to deformation of one or both components in the binary (Mowlavi et al. 2017). The parameters of the geometrical models are published in the catalogue, and an analysis of the results is presented in Mowlavi et al. (2022). A global ranking is also provided, which ranges from 0.4 to about 0.8 and sorts the candidates from least good to best candidates relative to their model fits. This catalogue of eclipsing binaries is purely based on the brightness variability and is complementary to the detailed non-single source analysis published in the different DR3 tables, e.g., `nss_two_body_orbit` (see for example Gaia Collaboration, Arenou et al. 2022).

---

[4] Few star classification were corrected since the publication of the archive table. The latter table gives 15,021 Cepheids
[5] https://www.cosmos.esa.int/web/gaia/iow_20220527.

### 3.4.5. Exoplanetary transits

The photometric precision of Gaia is high and stable enough to detect exoplanetary transits. The first confirmed transiting exoplanet of Gaia, Gaia-1b, was announced in March 2021 as a Gaia Image of the Week[6]. It was detected using the combined $G$, $G_{BP}$, $G_{RP}$ photometric measurements of Gaia, see Panahi et al. (2022b). Spectroscopic observations were conducted by the Large Binocular Telescope in Arizona to confirm its nature. In the article presenting the method to select candidates (Panahi et al. 2022b), a second confirmed case is also announced, Gaia-2b.

The DR3 table `vari_planetary_transit` presents 214 sources: 173 are known systems and 41 are new candidates (of which two were confirmed now, Gaia-1b and Gaia-2b). Very probably, the 39 remaining candidates contain some other actual planetary systems, which are just awaiting RV confirmation.

The exoplanetary transit search is applying a Box-Least-Square method (Panahi & Zucker 2021) on a first selection. This initial set consisted of 18 383 sources which came from the two possible paths of Special Variability Detection and Classification (see Fig. 3).

The table `vari_planetary_transit` contains information about the period and properties of the transit as derived from the Box-Least-Square for each source.

### 3.4.6. Long-period variables

The first Gaia catalogue of long-period variable stars was published in DR2 with 151 761 candidates with trimmed ranges in the $G$ magnitude larger than 0.2 mag variability (Mowlavi et al. 2018). In DR3, we present the second Gaia catalogue of long-period variable stars, in `vari_long_period_variable`, containing 1 720 558 sources with trimmed ranges in the $G$ magnitude larger than 0.1 mag amplitude (see Lebzelter et al. 2022). Further selection criteria include filters on such parameters as a combination of colour and brightness, on the number of epochs, and on $G$-band signal-to-noise. Periods and associated model amplitudes from the $G$-band light curves are provided for 392 240 sources. In addition, a flag based on low-resolution RP spectra is provided to identify carbon star candidates[7]. Some results are shown in Lebzelter et al. (2022), such as the period-luminosity relations for long-period variables in the solar neighbourhood and in Local Group galaxies.

### 3.4.7. Main-sequence oscillators

Using the GVD path, main-sequence oscillator candidate stars are not only available in the classification table (see above) but also in `vari_ms_oscillator`, with details about the best fre-quency and its harmonics (see Gaia Collaboration, De Ridder et al. 2022 for more details).

---

[6] https://www.cosmos.esa.int/web/gaia/iow_20210330
[7] For an illustration of the method, see the *Gaia* image of the week https://www.cosmos.esa.int/web/gaia/iow_20181115.

### 3.4.8. Microlensing

The microlensing events are being detected in real-time by the Gaia Science Alert system (cf. Hodgkin et al. 2021). This led, for example, to the discoveries such as the Gaia16aye event, which was fully resolved thanks to its complexity (binary lens) and to intense ground-based follow-ups (Wyrzykowski et al. 2020). However, a detailed a posteriori analysis of all the Gaia data searching for microlensing effect gives a different perspective and can lead to new discoveries.

Microlensing candidates were selected from a preliminary "extractor" method and confirmed in the SOS package, which fitted the Paczynski model without and with blending, see Wyrzykowski et al. (2022).

In the archive table `vari_ms_microlensing`, there are 363 reported events, among which 90 events were never reported before and were not discovered by other surveys.

Gaia microlensing events are unique as Gaia can provide both their photometric and astrometric time series. Analysis of such combined data sets in the future will lead to derive the nature of the lensing objects and might lead to the discovery of isolated stellar remnants such as neutron stars or black holes. A preliminary demonstration of Gaia capabilities was given in a Gaia Image of the Week[8].

### 3.4.9. Rotation modulation

Building on the DR2 experience, see Lanzafame et al. (2018), `vari_rotation_modulation` includes candidate stars of rotation modulation. Similarly as in DR2, the work package makes use of the SVD path for variability detection, identifying potential sources with rotational modulation and/or flares, with a refined analysis in the SOS pipeline to validate or not their status as stars displaying rotational modulation. For DR3, the pipeline was updated, and the analysis was modified to account for the longer photometric time series, allowing better detection of rotational periods, and for the improved outlier removal, together with different pipeline parameters, in particular a wider band in the colour-magnitude diagram near the main sequence. The details can be found in Distefano et al. (2022). In addition, a method was developed to identify spurious signals.

### 3.4.10. Short timescales

In the SVD path, further SOS tables were produced. The table `vari_short_timescale` provides 471 679 candidates showing short time scale phenomena (< below 0.5-1 day), together with timescale parameters. The same approach as in DR2 was used, therefore, we refer to Roelens et al. (2017) and Roelens et al. (2018) for details. The colour-magnitude diagram, Fig. 5, shows magnitude effects due to the change of windowing schemes or gating system. They happen at magnitudes 10.5, 12 and 13.2. Such observing strategy techniques still leave effects in the calibrations and increase the noise. This type of variability is spurious.

---

[8] https://www.cosmos.esa.int/web/gaia/iow_20210924.

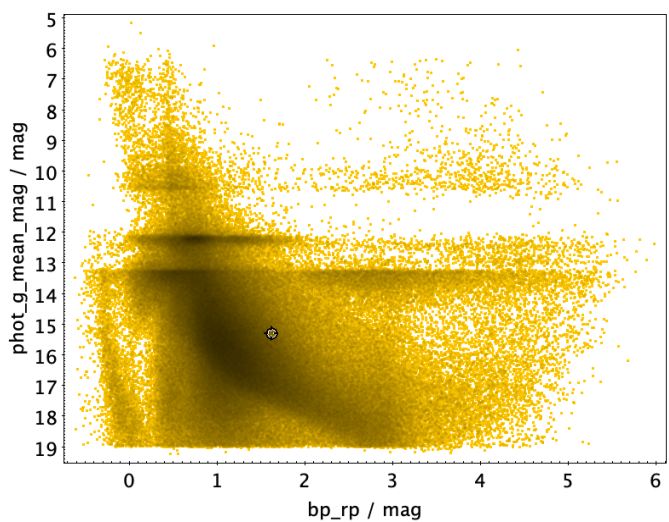

**Fig. 5.** Colour - (apparent) magnitude diagram for the short time scale candidates. We remark that there are remaining artefacts. Care should be taken in using this sample.

## 3.5. Verification, validation, and overlaps

The variability analysis underwent several verification/validation steps toward the final publication in DR3. As part of operational runs, so-called 'violation rules' were implemented to omit the values of output parameters beyond the expected ranges. For example, negative errors or periods/frequencies are not allowed. In some cases, complex rules were implemented which depended on other parameters. Such rules allowed the early detection of software bugs and their correction for the final operational run.

Before exporting our data for ingestion in the pre-DR3 archive, different teams validated their samples of candidate variable objects based on different criteria described in detail in the relevant papers. Either automatic, reproducible criteria were used, or visual inspections were done.

A specific validation was done for the YSO variability type provided by classification, which detected 79 375 of such young stellar objects thanks to their variability. Cross-matches with several catalogues were used to assess completeness and contamination. Most of the classified YSOs are located in the directions of known star forming regions and in the Galactic mid-plane. Despite a percent level of completeness, about 40 000 new YSO candidates were found. More details can be found in (Marton et al. 2022).

In the next step before export, we carefully investigated overlaps of sources between SOS packages. Initial counts identified large overlaps between different SOS candidates and short timescale candidates (more than 300k sources) and eclipsing binaries (about 260k). However, overlapping numbers were significantly lower ($< 1-2$k) for other classes. We devised overlapping rules to attribute sources in one SOS package (e.g., if in both AGN and microlensing – or rotation modulation or short timescale – list of candidates, the source was attributed only to AGN; inversely, for example, if a source was identified as AGN and Cepheid, it was eventually attributed to Cepheid), or to allow overlap (e.g., short timescale and LPV - or EB or MS oscillator or rotation modulation - candidates). The overlaps for the SOS classes can be found in Fig. 6.

We emphasise that the best class entry in `vari_classifier_result` does not necessarily mean that a source will only appear in the related SOS table. However, for the vast majority of cases of SOS packages following classification, this is the case. For the other cases, while some packages follow the classification step, sources can be identified through other paths (e.g., SVD and microlensing extractor). SOS packages in the GVD path take a range of classification probabilities as input sources, i.e., the best class only was not necessarily considered. So, one can find, e.g., a source with "SOLAR_LIKE" best class and this source also in `vari_short_timescale`, or an "ECL" best class and an entry in `vari_eclipsing_binary`.

Eventually, checks were performed when exporting our data to DPAC and inclusion in the pre-releases of the DR3 archive. We note that further sources were removed at the DPAC level before eventual publication in DR3.

## 4. Variability types and their properties

There are two origins of variable sources in the Gaia archive. For a variability type, two table types (if they exist) should be checked: classification and specific object studies.

In Table 2, we provide the total numbers per variability type and their origin (either from classification or specific object studies). We also note the counts from one channel only (e.g. from classification only, with no counterpart in specific object studies), and we compute as well the overlaps between the same classes coming either from classification or from the table of the specific object studies. The denomination of the `best_class_name` and for the table name are given as they appear in the archive. In general, the sets formed by sources for a given variability type from classification should have higher completeness, and the sources from specific studies should have lower contamination. There are exceptions to this. For example, the set of eclipsing binaries from classification is slightly smaller. The set of exoplanetary transits from classification was limited to match the final list of Specific Object Studies. We give the detailed overlaps between the classification and SOS packages in Fig. 7. We remark that the short timescale and S types seem to cover very different populations.

The `vari_summary` table provides a wide range of statistics and can be used for statistical descriptions of the variability types. Tables 3 and 4 present properties of these variability types, as they appear either from the supervised classification or from the specific object studies. These tables allow us to determine the distribution of parameters for each variability class. We provide here the first and last decil and the median of these distributions for the trimmed range in $G$ mag, the ratio of the colour variation (i.e. ratio of trimmed ranges of $G_{BP}$ over $G_{RP}$), and depending on the table, the colour and absolute magnitude distribution or the period.

Some joins of DR3 tables between the type groups from classification and the main sequence oscillators allow us to split some type groups. As an example, we are able to derive properties of $\gamma$ Doradus stars and $\delta$ Scuti stars, see Table 5. However, it should be noted that the period distribution of the main sequence oscillators contains gaps and therefore bias somewhat these properties.

**A warning on the determination of the sample statistics weighting** There are different methods to estimate statistical values of a distribution for a given sample. As different estimators are used, there are, therefore, different possible values. Moreover, some estimators can be less precise, less robust to outlying values, or some might present biases under certain conditions. Here we show a difference in the archive tables for the sample mean, one of the most basic statistical quantities.

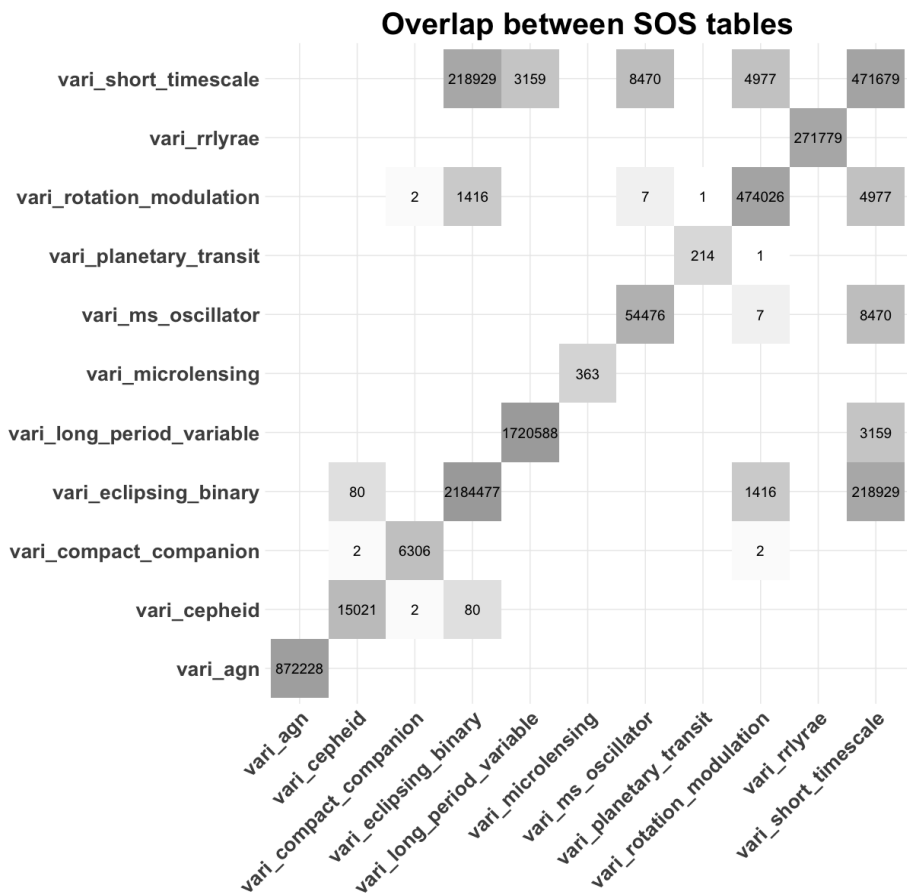

**Fig. 6.** Overlap between the variability types from different SOS tables.

We are comparing the value `phot_g_mean_mag` of the table `gaia_source` and the value `mean_mag_g_fov` of the table `vari_summary`. The first value is computed as a weighted mean on the CCD fluxes and then converted into magnitude, while the second is computed as a mean on the FoV magnitudes (without a weighting procedure).

Differences are expected as the means are computed for the first case before a logarithmic transformation and in the second case after. Another source of possible differences is that the data sets that have been included in computing the estimated value may not be strictly similar because some observations might have been rejected (e.g., see Sect. 3.1).

Here we want to point out another difference. The weighting procedure may bias the results when the signal is highly variable. This bias emerges because the uncertainty is correlated with the value and therefore is skewing the result (Lecoeur-Taibi & Eyer 2016). We use large-amplitude variables such as long-period variables, Cepheids, RR Lyrae stars, and eclipsing binaries to demonstrate this. In Fig. 8, we plot the difference (weighted-unweighted) of the means as a function of the unweighted mean. The flux-weighted means provide fainter values in general than the unweighted, the larger the amplitude, the larger the bias is. Details will be presented in a separate article (Eyer et al. 2022). This section serves as a warning and a demonstration that extreme care should be taken even with simple statistics.

## 5. Completeness and contamination

Given the large amount and variety of classes of variable stars, when classifying them, it is important to provide estimates of completeness and contamination[9]. We give these estimates in Table 6 for the SOS tables. For the completeness and contamination estimates of the variability types from the classifiers, we refer to the article dedicated to the classification Rimoldini et al. (2022a). We remark that these estimates depend strongly on the variability type. For some variability types, we did not determine these estimates either because the current knowledge is too shallow (compact companions) or because the variability class is too broad (short timescale variability, main sequence oscillators).

## 6. HR diagrams of variable stars

Here we show the position of variable stars in Hertzsprung-Russell (HR)/colour-magnitude diagrams for different stellar systems: the LMC, the SMC (cf. Figs. A.2 and 10). For different types of variable stars in our Galaxy, the HR diagrams can be found in Rimoldini et al. (2022a), As these systems have different chemical compositions, ages and formation history, their variable star content is different. Furthermore, the observational constraints are different, and as a consequence, the possible contaminations are also different. For the SMC and LMC, we assume that they are at a fixed distance, and we do not correct any depth effect so that plotting the apparent magnitude is directly related to their absolute magnitude. We also make a selection on the parallax that should be smaller than 5 times its error and on proper motions following Gaia Collaboration et al. (2021).

---

[9] In some of the DR3 articles, the term purity (1−contamination) is sometimes preferred

**Table 2.** Total numbers of sources per variability type. The counts are also given for the intersection of classification set and Specific Object Studies set for the same variability type. The detailed overlaps between different classes from Classification and Specific Object Studies are in Fig. 7. In parenthesis are the numbers, for the same variability type, of the sources of classification/SOS at the exclusion of SOS/classification respectively.

| Variability type / type group | Total | Classification name Count (Only Classif.) | SOS table name Count (Only SOS) | Classif. ∩ SOS |
|---|---|---|---|---|
| $\alpha^2$ CVn and associated stars | 10 779 | ACV\|...\|SXARI 10 779 | - | - |
| $\alpha$ Cygni stars | 329 | ACYG 329 | - | - |
| Active Galactic Nuclei (QSO) | 1 035 254 | AGN 1 035 207 (163 026) | vari_agn 872 228 (47) | 872 181 |
| $\beta$ Cephei stars | 1475 | BCEP 1475 | - | - |
| Be stars, $\gamma$ Cas and associated stars | 8560 | BE\|...\|WR 8560 | - | - |
| Cepheids | 16 175 | CEP 16 141 (1 154) | vari_cepheid 15 021 (34) | 14 987 |
| Cataclysmic variables | 7306 | CV 7306 | - | - |
| $\delta$ Scuti/$\gamma$ Doradus/SX Phoenicis stars | 748 058 | DSCT/GDOR 748 058 | | |
| Eclipsing binaries | 2 184 496 | ECL 2 184 356 (19) | vari_eclipsing_binary 2 184 477 (140) | 2 184 337 |
| Ellispoidal variations | 65 300 | ELL 65 300 | - | - |
| Exoplanetary transits | 214 | EP 214 (0) | vari_planetary_transit 214 (0) | 214 |
| Long-period variables | 2 326 297 | LPV 2 325 775 (605 709) | vari_long_period_variable 1 720 588 (522) | 1 720 066 |
| Microlensing events | 430 | MICROLENSING 254 (67) | vari_microlensing 363 (176) | 187 |
| R Coronae Borealis stars | 153 | RCB 153 | - | - |
| RR Lyrae stars | 297 981 | RR 297 778 (26 202) | vari_rrlyrae 271 779 (203) | 271 576 |
| RS CVn | 742 263 | RS 742 263 | - | - |
| SDB | 893 | SDB 893 | - | - |
| Short time scale | 983 185 | S 512 005 (511 506) | vari_short_timescale 471 679 (471 180) | 499 |
| Supernovae | 3029 | SN 3029 | - | - |
| Solar-like variability | 2 306 297 | SOLAR_LIKE 1 934 844 (1 832 271) | vari_rotation_modulation 474 026 (371 453) | 102 573 |
| Slowly Pulsating B star | 1228 | SPB 1228 | - | - |
| Symbiotic System | 649 | SYST 649 | - | - |
| Variable White Dwarfs | 910 | WD 910 | - | - |
| Young Stellar Objects | 79 375 | YSO 79 375 | - | - |

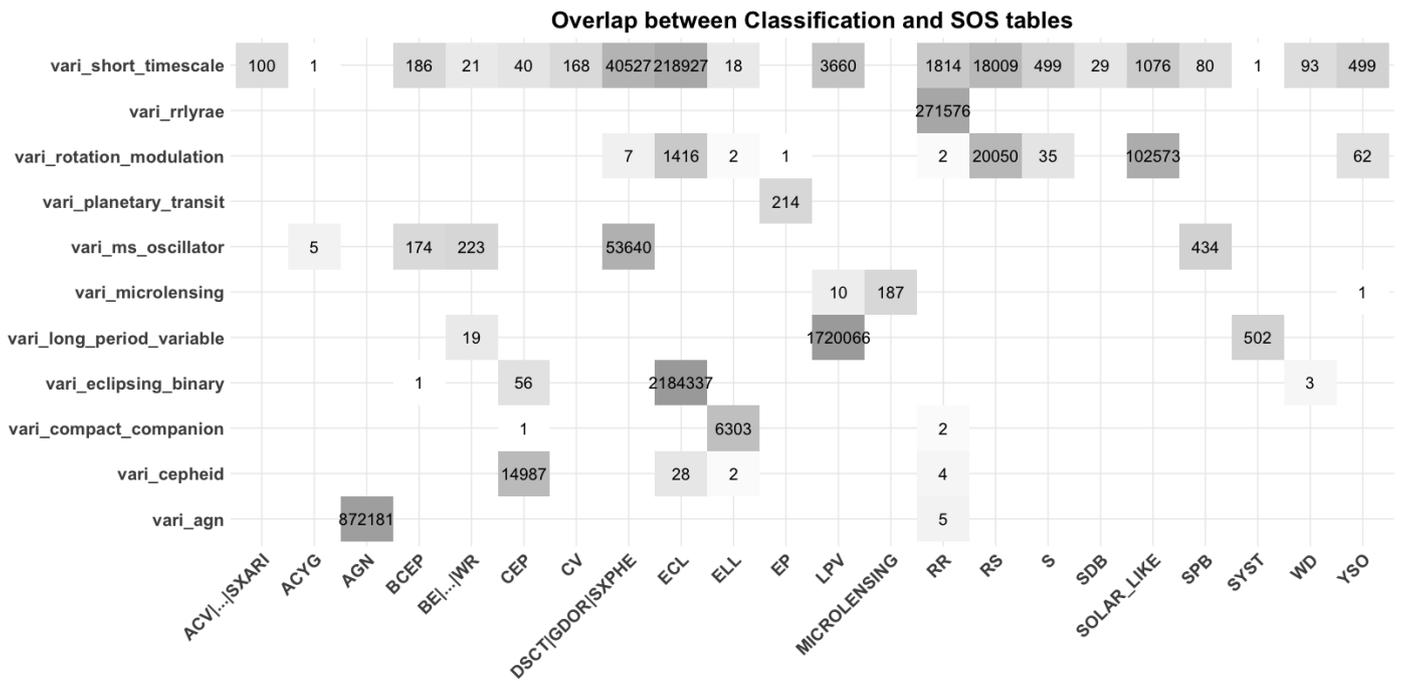

**Fig. 7.** Overlap between `best_class_name` in `vari_classifier_result` and SOS tables.

There are several interesting observations. For example, the location of Cepheids in the LMC and the SMC is not just a translation about half a magnitude fainter for the SMC. There is an intrinsic change that comes partly from the different metallicity. Stars of lower metallicities have blue loops at lower masses, and therefore the SMC will have a larger fraction of Cepheids compared to its population.

## 7. Gaia-TESS synergy for exoplanets

The TESS mission is surveying bright stars in search of exoplanets using the transit method (Ricker et al. 2015). TESS issues a monthly list of newly discovered candidates of transiting exoplanets. However, the TESS point spread function is large (about 1 arcmin), and therefore, the light of each target star blends with the light from nearby sources. Therefore, follow-up photometric observations are required in order to exclude false detections of transits that are actually caused by blending with a nearby eclipsing binary. Thanks to the high spatial resolution and its multi-epoch photometry, Gaia can help determine the true nature of the candidate variability. A collaboration between Gaia and TESS was established, and TESS exoplanet candidates are regularly analysed within the Gaia consortium using unpublished Gaia data, and the results shared to the TESS collaboration. With a rate of about 5%, Gaia can identify false-positive candidates, and at a rate of about 5%, Gaia can even confirm true detections of transiting planets. Panahi et al. (2022a) present the details of this ongoing cooperation.

## 8. Conclusions

The Gaia DR3 catalogues of variable sources will serve as a basis for diverse scientific studies. This article summarises the processing and analysis that we did for DR3. More details can be found in the Gaia DR3 articles on specific topics, as listed in Sect. 3. One of these articles was written in the context of a Performance Verification paper (Gaia Collaboration, De Ridder et al. 2022) and thus allowed for a deeper analysis of the pulsations in main-sequence OBAF-type stars.

The future data releases DR4 and DR5 will bring major improvements:

- at each of these data releases, the number of measurements will roughly double with respect to the previous one;
- there will be an improvement in calibrations and the techniques to detect and/or avoid instrumental effects;
- the use of the BP, RP and RVS spectra time series, radial velocities, and per-CCD time series will be considered for each of the variability types when deemed useful;
- classification training sets will grow in quality, and the attributes of the classification will be further tailored for each variability class;
- an unsupervised classification will be implemented in addition to the supervised classification.

For some sky regions, the total number of per-CCD measurements may be above 2000 for a 10-year mission. The increase in time span and the number of measurements available will open the window to detect multi-periodic signals with a high-frequency resolution (provided that the star behaviour is stable over ten years).

There will be even more synergy in the future with other large surveys such as ZTF (Graham et al. 2019), TESS (Ricker et al. 2015), LSST (LSST Science Collaboration et al. 2009), PLATO (Rauer et al. 2014), etc. As mentioned in Sect. 7, one identified synergy which is currently under exploitation takes advantage of the much higher spatial resolution of Gaia to identify photometric blends in TESS to avoid the false detection of exoplanetary transits.

Combining ZTF and/or LSST data with Gaia will be very beneficial as the sampling strategies are very different while the photometric precisions and depths have some overlaps, thus allowing for combined data analyses.

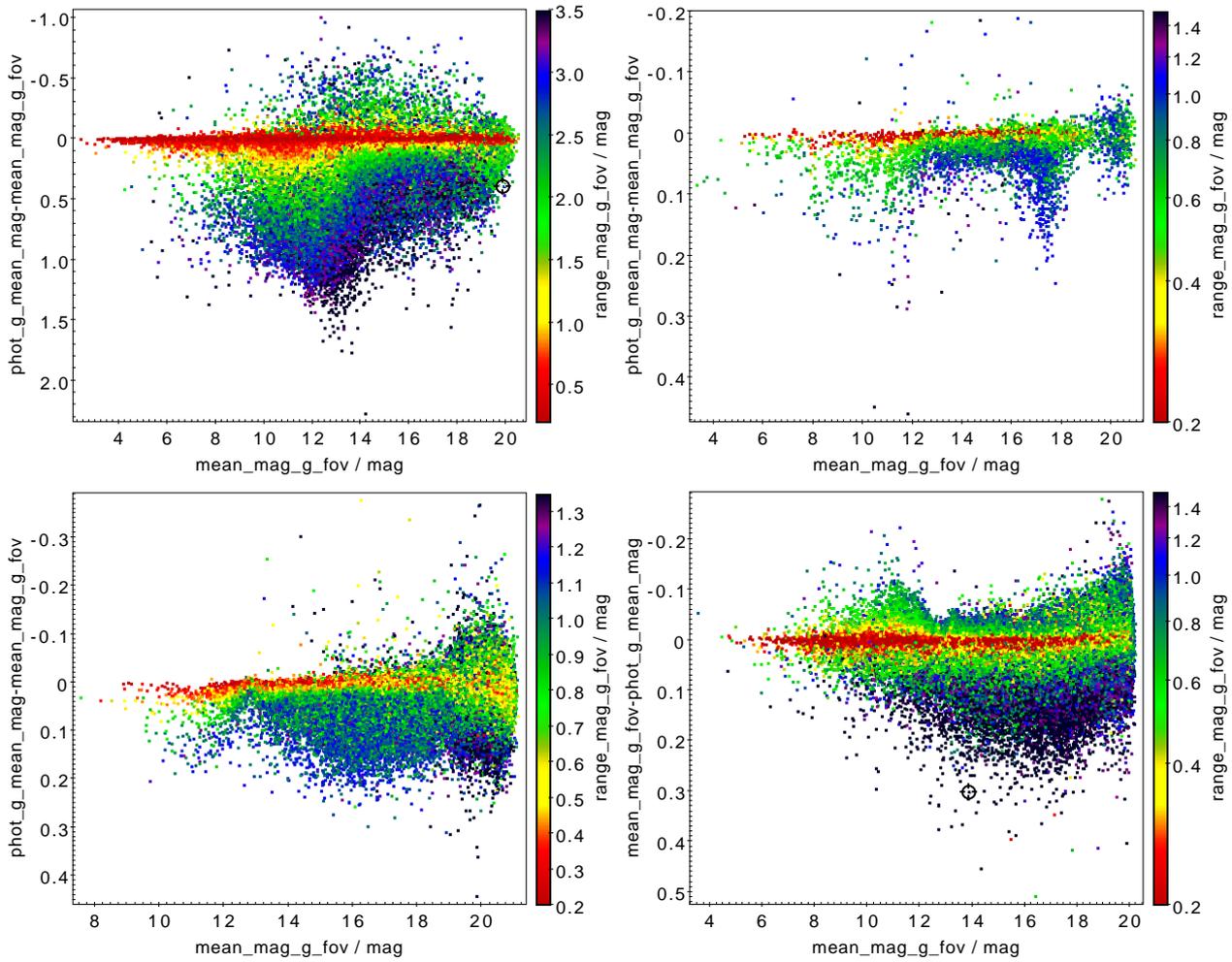

**Fig. 8.** Diagrams of the differences between two estimators of the mean found in the Gaia archive (`mean_mag_g_fov` in the `vari_summary` table and `phot_g_mean_mag` in the `gaia_source` table). The weighted mean on fluxes versus the unweighted mean on FoV magnitudes. The colour scale is the range of the magnitude distribution, so as function of the signal amplitude. Top left panel: for long-period variables. Top right: Cepheids. Bottom left: RR Lyrae stars. Bottom right: eclipsing binaries.

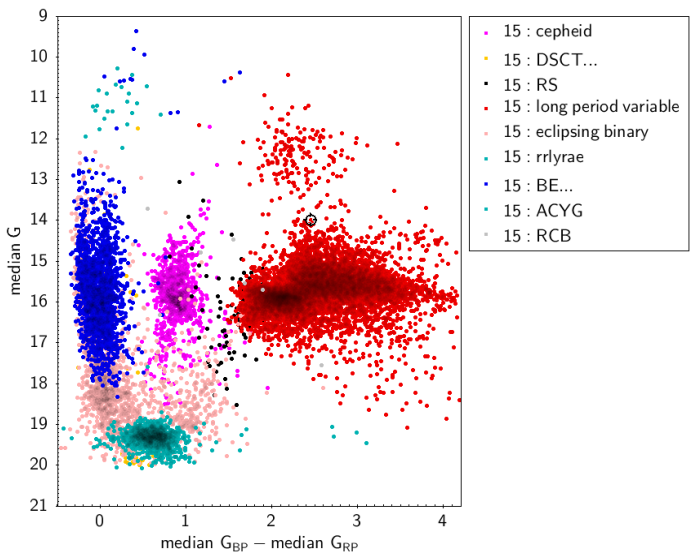

**Fig. 9.** Colour-Magnitude diagram for the variables in the LMC.

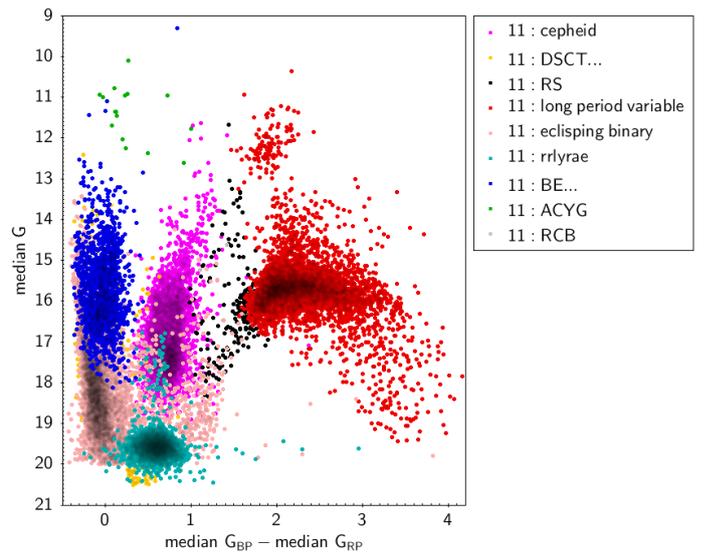

**Fig. 10.** Colour-Magnitude diagram for the variables in the SMC.

*Acknowledgements.* This work presents results from the European Space Agency (ESA) space mission Gaia. Gaia data are being processed by the Gaia Data Processing and Analysis Consortium (DPAC). Funding for the

**Table 3.** Variability Types from Classification together with variability quantities in photometry and in the colour-magnitude diagram.

| Variability type | Number[1] | TR($G$)[2] (mag) Q10/Q50/Q90 | TR($G_{BP}$)/TR($G_{RP}$)[3] Q10/Q50/Q90 | $M_G$[4] (mag) Q10/Q50/Q90 | $G_{BP}$-$G_{RP}$[4] (mag) Q10/Q50/Q90 |
|---|---|---|---|---|---|
| ACV\|CP\|...\|SXARI[5] | 10 779 | 0.018/0.025/0.064 | 0.79/1.31/1.96 | −0.63/0.88/2.06 | −0.016/0.22/0.45 |
| ACYG | 329 | 0.031/0.053/0.084 | 0.59/0.97/1.23 | −6.06/−4.86/−3.64 | 0.28/0.65/1.17 |
| AGN | 1 035 207 | 0.22/0.36/0.58 | 0.75/1.16/1.84 | NA | 0.39/1.22/2.47 |
| BCEP | 1475 | 0.020/0.031/0.066 | 0.95/1.21/1.56 | −2.74/−1.37/−0.33 | −0.032/0.34/1.13 |
| BE\|GCAS\|SDOR\|WR | 8560 | 0.047/0.098/0.26 | 0.52/0.79/1.24 | −2.83/−1.53/−0.37 | 0.16/0.78/2.16 |
| CEP | 16 141 | 0.22/0.46/0.84 | 1.25/1.59/1.83 | −2.83/−1.02/1.14 | 0.99/1.76/2.96 |
| CV | 7306 | 0.64/1.7/3.6 | 0.84/1.22/1.73 | 5.12/8.39/11.02 | 0.23/0.66/1.26 |
| DSCT\|GDOR\|SXPHE | 748 058 | 0.017/0.024/0.065 | 0.89/1.35/1.85 | 1.46/2.52/3.32 | 0.46/0.62/0.85 |
| ECL | 2 184 356 | 0.11/0.28/0.57 | 0.93/1.28/2.32 | 2.27/4.27/6.61 | 0.73/1.19/1.84 |
| ELL | 65 300 | 0.040/0.076/0.15 | 1.16/2.10/4.51 | 1.23/3.31/5.71 | 1.30/1.78/2.66 |
| EP | 214 | 0.009/0.016/0.029 | 0.9§/1.29/1.72 | 3.05/4.23/5.86 | 0.65/0.84/1.21 |
| LPV[6] | 2 352 775 | 0.11/0.19/0.58 | 1.62/2.64/7.33 | −2.33/−1.23/0.91 | 2.32/3.34/4.80 |
| MICROLENSING | 254 | 0.22/0.62/1.46 | 0.86/1.22/2.52 | NA | 1.43/1.79/2.70 |
| RCB | 153 | 1.18/4.28/6.73 | 0.84/1.14/1.38 | −3.34/−1.27/3.31 | 1.42/2.15/3.11 |
| RR | 297 778 | 0.31/0.53/0.89 | 1.10/1.54/1.93 | 0.19/0.02/3.70 | 0.48/0.75/1.40 |
| RS | 742 263 | 0.054/0.094/0.17 | 1.10/1.48/2.27 | 2.20/4.62/6.62 | 1.02/1.32/1.69 |
| S | 512 005 | 0.34/0.58/0.81 | 1.17/1.88/3.19 | 5.47/7.76/10.1 | 1.09/1.47/1.91 |
| SDB | 893 | 0.025/0.033/0.044 | 0.54/0.90/1.54 | 3.90/4.40/4.83 | −0.42/−0.34/−0.22 |
| SN | 3029 | 0.62/2.53/4.30 | 0.63/1.09/1.69 | NA | NA |
| SOLAR_LIKE | 1 934 844 | 0.019/0.028/0.056 | 1.22/1.68/3.04 | 4.79/6.01/8.29 | 0.96/1.26/2.04 |
| SPB | 1228 | 0.029/0.038/0.058 | 0.78/1.17/1.45 | −1.51/−0.19/0.66 | −0.17/−0.052/0.042 |
| SYST | 649 | 0.17/0.32/0.77 | 1.11/2.41/3.84 | −2.88/−1.39/1.08 | 1.40/2.65/3.70 |
| WD | 910 | 0.035/0.071/0.016 | 0.51/0.83/1.39 | 7.90/11.5/12.1 | −0.46/−0.007/0.18 |
| YSO | 79 375 | 0.037/0.085/0.21 | 1.53/3.84/9.14 | 5.13/7.86/9.76 | 1.91/2.74/3.46 |

**Notes.** (1) Total number of sources in `vari_classifier_result`. (2) Trimmed range in $G$, using sources with $N > 15$ epochs in $G_{BP}$ and $G_{RP}$. (3) Calculated using the ratio of the trimmed ranges in $G_{BP}$ and $G_{RP}$, for sources with $N > 15$ epochs in $G_{BP}$ and $G_{RP}$ and median magnitudes $< 20$ in $G_{BP}$ and $< 19.5$ in $G_{RP}$. (4) Calculated using a `gaia_source.parallax_over_error > 5`, and for sources with $N > 15$ epochs in $G_{BP}$ and $G_{RP}$. (5) ACV\|CP\|MCP\|ROAM\|ROAP\|SXARI. (6) The lower boundary for LPV was set to 0.1 mag, see Lebzelter et al. (2022).

DPAC is provided by national institutions, in particular the institutions participating in the Gaia MultiLateral Agreement (MLA). The Gaia mission website is https://www.cosmos.esa.int/gaia. The Gaia archive website is https://archives.esac.esa.int/gaia. The Gaia mission and data processing have financially been supported by, in alphabetical order by country:the Algerian Centre de Recherche en Astronomie, Astrophysique et Géophysique of Bouzareah Observatory; the Austrian Fonds zur Förderung der wissenschaftlichen Forschung (FWF) Hertha Firnberg Programme through grants T359, P20046, and P23737; the BELgian federal Science Policy Office (BELSPO) through various PROgramme de Développement d'Expériences scientifiques (PRODEX) grants and the Polish Academy of Sciences - Fonds Wetenschappelijk Onderzoek through grant VS.091.16N, and the Fonds de la Recherche Scientifique (FNRS), and the Research Council of Katholieke Universiteit (KU) Leuven through grant C16/18/005 (Pushing AsteRoseismology to the next level with TESS, GaiA, and the Sloan DIgital Sky SurvEy – PARADISE); the Brazil-France exchange programmes Fundação de Amparo à Pesquisa do Estado de São Paulo (FAPESP) and Coordenação de Aperfeicoamento de Pessoal de Nível Superior (CAPES) - Comité Français d'Evaluation de la Coopération Universitaire et Scientifique avec le Brésil (COFECUB); the Chilean Agencia Nacional de Investigación y Desarrollo (ANID) through Fondo Nacional de Desarrollo Científico y Tecnológico (FONDECYT) Regular Project 1210992 (L. Chemin); the National Natural Science Foundation of China (NSFC) through grants 11573054, 11703065, and 12173069, the China Scholarship Council through grant 201806040200, and the Natural Science Foundation of Shanghai through grant 21ZR1474100; the Tenure Track Pilot Programme of the Croatian Science Foundation and the École Polytechnique Fédérale de Lausanne and the project TTP-2018-07-1171 'Mining the Variable Sky', with the funds of the Croatian-Swiss Research Programme; the Czech-Republic Ministry of Education, Youth, and Sports through grant LG 15010 and INTER-EXCELLENCE grant LTAUSA18093, and the Czech Space Office through ESA PECS contract 98058; the Danish Ministry of Science; the Estonian Ministry of Education and Research through grant IUT40-1; the European Commission's Sixth Framework Programme through the European Leadership in Space Astrometry (ELSA) Marie Curie Research Training Network (MRTN-CT-2006-033481), through Marie Curie project PIOF-GA-2009-255267 (Space AsteroSeismology & RR Lyrae stars, SAS-RRL), and through a Marie Curie Transfer-of-Knowledge (ToK) fellowship (MTKD-CT-2004-014188); the European Commission's Seventh Framework Programme through grant FP7-606740 (FP7-SPACE-2013-1) for the Gaia European Network for Improved data User Services (GENIUS) and through grant 264895 for the Gaia Research for European Astronomy Training (GREAT-ITN) network; the European Cooperation in Science and Technology (COST) through COST Action CA18104 'Revealing the Milky Way with Gaia (MW-Gaia)'; the European Research Council (ERC) through grants 320360, 647208, and 834148 and through the European Union's Horizon 2020 research and innovation and excellent science programmes through Marie Skłodowska-Curie grant 745617 (Our Galaxy at full HD – Gal-HD) and 895174 (The build-up and fate of self-gravitating systems in the Universe) as well as grants 687378 (Small Bodies: Near and Far), 682115 (Using the Magellanic Clouds to Understand the Interaction of Galaxies), 695099 (A sub-percent distance scale from binaries and Cepheids – CepBin), 716155 (Structured ACCREtion Disks – SACCRED), 951549 (Sub-percent calibration of the extragalactic distance scale in the era of big surveys – UniverScale), and 101004214 (Innovative Scientific Data Exploration and Exploitation Applications for Space Sciences – EXPLORE); the European Science Foundation (ESF), in the framework of the Gaia Research for European Astronomy Training Research Network Programme (GREAT-ESF); the European Space Agency (ESA) in the framework of the Gaia project, through the Plan for European Cooperating States (PECS) programme through contracts C98090 and 4000106398/12/NL/KML for Hungary, through contract 4000115263/15/NL/IB for Germany, and through PROgramme de Développement d'Expériences scientifiques (PRODEX) grant 4000127986 for Slovenia; the Academy of Finland through grants 299543, 307157, 325805, 328654, 336546, and 345115 and the Magnus Ehrnrooth Foundation; the French Centre National d'Études Spatiales (CNES), the Agence Nationale de la Recherche (ANR) through grant ANR-10-IDEX-0001-02 for the 'Investissements d'avenir' programme, through grant ANR-15-CE31-0007 for project 'Modelling the Milky Way in the Gaia era' (MOD4Gaia), through grant ANR-14-CE33-0014-01 for project 'The Milky Way disc formation in the Gaia era' (ARCHEOGAL), through grant ANR-15-CE31-0012-01 for project 'Unlocking the potential of Cepheids as primary distance calibrators' (UnlockCepheids), through grant ANR-19-CE31-0017 for project 'Secular evolution of galxies' (SEGAL), and through grant ANR-18-CE31-0006 for project 'Galactic Dark Matter' (GaDaMa), the Centre National de la Recherche Scientifique (CNRS) and its SNO Gaia of the Institut des Sciences de l'Univers (INSU), its Programmes Nationaux: Cosmologie et Galaxies (PNCG), Gravitation Références Astronomie Métrologie (PNGRAM), Planétologie (PNP), Physique et Chimie du Milieu Interstellaire (PCMI), and Physique Stellaire (PNPS), the 'Action Fédératrice Gaia' of the Observatoire de Paris, the Région de Franche-Comté, the Institut National Polytechnique (INP) and the Institut National de Physique nucléaire et de Physique des Particules (IN2P3) co-funded by CNES; the German Aerospace Agency (Deutsches Zentrum für Luft- und Raumfahrt e.V., DLR) through grants 50QG0501, 50QG0601, 50QG0602, 50QG0701, 50QG0901, 50QG1001, 50QG1101,

**Table 4.** Variability Types from Specific Objects Studies together with variability quantities in photometry and in time.

| Variability type[1] | Number | TR($G$)[2] (mag) Q10/Q50/Q90 | TR($G_{BP}$)/TR($G_{RP}$)[3] Q10/Q50/Q90 | Period/Time scale Q01/Q50/Q99 |
|---|---|---|---|---|
| AGN | 872 228 | 0.21/0.35/0.58 | 0.75/1.16/1.82 | |
| Compact Companions | 6 306 | 0.14/0.16/0.19 | 1.09/1.64/2.81 | 0.28/0.41/0.89 |
| Cepheids | 15 021 | 0.22/0.46/0.84 | 1.26/1.59/1.82 | 1.49/3.90/16.5 |
| – $\delta$ Cep | 12 897 | 0.22/0.45/0.81 | 1.31/1.60/1.83 | 1.55/3.76/15.0 |
| – T2CEP | 1 534 | 0.19/0.55/0.96 | 1.14/1.42/1.79 | 1.46/8.18/30.0 |
| – ACEP | 376 | 0.29/0.60/0.96 | 1.17/1.52/1.78 | 0.84/1.28/1.89 |
| Eclipsing binaries | 2 184 477 | 0.11/0.28/0.57 | 0.93/1.28/2.32 | 0.28/0.48/3.83 |
| Long-period variables | 1 720 588 | 0.11/0.19/0.51 | 1.66/2.56/7.07 | 53.1/245.8/565.6 |
| Microlensing events | 363 | 0.14/0.52/1.34 | 0.92/1.22/2.40 | 25.3/59.3/161.2 |
| Main-Sequence Oscillators | 54 476 | 0.022/0.044/0.177 | 0.81/1.35/1.74 | 0.048/0.087/2.53 |
| Planetary transits | 214 | 0.0094/0.016/0.029 | 0.91/1.29/1.72 | 0.59/1.26/4.88 |
| Rotation Modulation | 474 026 | 0.013/0.032/0.099 | 1.22/1.78/3.51 | 0.41/2.24/8.44 |
| RR Lyrae stars | 272 428 | 0.32/0.54/0.90 | 1.09/1.53/1.91 | 0.47/0.57/0.67 |
| – RRab | 140 597 | 0.36/0.67/0.94 | 1.13/1.53/1.90 | 0.47/0.57/0.67 |
| – RRc | 63 010 | 0.28/0.38/0.49 | 1.01/1.54/1.95 | 0.26/0.32/0.38 |
| – RRd | 1930 | 0.43/0.51/0.66 | 0.99/1.46/1.76 | 0.46/0.49/0.55 |
| Short time scale | 471 679 | 0.014/0.13/0.48 | 0.96/1.30/2.58 | 0.08/0.14/0.44 |

**Notes.** (1) Total number of sources in different SOS tables. (2) Trimmed range in $G$, using sources with $N > 15$ epochs in $G_{BP}$ and $G_{RP}$. (3) Calculated using the ratio of the trimmed ranges in $G_{BP}$ and $G_{RP}$, for sources with $N > 15$ epochs in $G_{BP}$ and $G_{RP}$ and median magnitudes $< 20$ in $G_{BP}$ and $< 19.5$ in $G_{RP}$.

**Table 5.** Variability Types from an intersection of Classification with Specific Objects Studies together with variability quantities in photometry and in time.

| Variability type[1] | Number | TR($G$)[2] (mag) Q10/Q50/Q90 | TR($G_{BP}$)/TR($G_{RP}$)[3] Q10/Q50/Q90 | Period/Time scale Q01/Q50/Q99 |
|---|---|---|---|---|
| $\delta$ Scuti[1] | 37 530 | 0.02/0.05/0.22 | 0.90/1.43/1.76 | 0.05/0.07/0.11 |
| $\gamma$ Doradus[1] | 16 110 | 0.02/0.04/0.07 | 0.69/1.15/1.65 | 0.58/1.66/5.35 |

**Notes.** (1) Total number of sources using a cut in period of 0.25 d to separate $\delta$ Scuti stars from $\gamma$ Doradus stars, the latter having longer periods. (2) Trimmed range in $G$, using sources with $N > 15$ epochs in $G_{BP}$ and $G_{RP}$. (3) Calculated using the ratio of the trimmed ranges in $G_{BP}$ and $G_{RP}$, for sources with $N > 15$ epochs in $G_{BP}$ and $G_{RP}$ and median magnitudes $< 20$ in $G_{BP}$ and $< 19.5$ in $G_{RP}$.

50QG1401, 50QG1402, 50QG1403, 50QG1404, 50QG1904, 50QG2101, 50QG2102, and 50QG2202, and the Centre for Information Services and High Performance Computing (ZIH) at the Technische Universität Dresden for generous allocations of computer time; the Hungarian Academy of Sciences through the Lendület Programme grants LP2014-17 and LP2018-7 and the Hungarian National Research, Development, and Innovation Office (NKFIH) through grant KKP-137523 ('SeismoLab'); the Science Foundation Ireland (SFI) through a Royal Society - SFI University Research Fellowship (M. Fraser); the Israel Ministry of Science and Technology through grant 3-18143 and the Tel Aviv University Center for Artificial Intelligence and Data Science (TAD) through a grant; the Agenzia Spaziale Italiana (ASI) through contracts I/037/08/0, I/058/10/0, 2014-025-R.0, 2014-025-R.1.2015, and 2018-24-HH.0 to the Italian Istituto Nazionale di Astrofisica (INAF), contract 2014-049-R.0/1/2 to INAF for the Space Science Data Centre (SSDC, formerly known as the ASI Science Data Center, ASDC), contracts I/008/10/0, 2013/030/I.0, 2013-030-I.0.1-2015, and 2016-17-I.0 to the Aerospace Logistics Technology Engineering Company (ALTEC S.p.A.), INAF, and the Italian Ministry of Education, University, and Research (Ministero dell'Istruzione, dell'Università e della Ricerca) through the Premiale project 'MIning The Cosmos Big Data and Innovative Italian Technology for Frontier Astrophysics and Cosmology' (MITiC); the Netherlands Organisation for Scientific Research (NWO) through grant NWO-M-614.061.414, through a VICI grant (A. Helmi), and through a Spinoza prize (A. Helmi), and the Netherlands Research School for Astronomy (NOVA); the Polish National Science Centre through HARMONIA grant 2018/30/M/ST9/00311 and DAINA grant 2017/27/L/ST9/03221 and the Ministry of Science and Higher Education (MNiSW) through DIR/WK/2018/12; the Portuguese Fundação para a Ciência e a Tecnologia (FCT) through national funds, grants SFRH/BD/128840/2017 and PTDC/FIS-AST/30389/2017, and work contract DL 57/2016/CP1364/CT0006, the Fundo Europeu de Desenvolvimento Regional (FEDER) through grant POCI-01-0145-FEDER-030389 and its Programa Operacional Competitividade e Internacionalização (COMPETE2020) through grants UIDB/04434/2020 and UIDP/04434/2020, and the Strategic Programme UIDB/00099/2020 for the Centro de Astrofísica e Gravitação (CENTRA); the Slovenian Research Agency through grant P1-0188; the Spanish Ministry of Economy (MINECO/FEDER, UE), the Spanish Ministry of Science and Innovation (MICIN), the Spanish Ministry of Education, Culture, and Sports, and the Spanish Government through grants BES-2016-078499, BES-2017-083126, BES-C-2017-0085, ESP2016-80079-C2-1-R, ESP2016-80079-C2-2-R, FPU16/03827, PDC2021-121059-C22, RTI2018-095076-B-C22, and TIN2015-65316-P ('Computación de Altas Prestaciones VII'), the Juan de la Cierva Incorporación Programme (FJCI-2015-2671 and IJC2019-04862-I for F. Anders), the Severo Ochoa Centre of Excellence Programme (SEV2015-0493), and MICIN/AEI/10.13039/501100011033 (and the European Union through European Regional Development Fund 'A way of making Europe') through grant RTI2018-095076-B-C21, the Institute of Cosmos Sciences University of Barcelona (ICCUB, Unidad de Excelencia 'María de Maeztu') through grant CEX2019-000918-M, the University of Barcelona's official doctoral programme for the development of an R+D+i project through an Ajuts de Personal Investigador en Formació (APIF) grant, the Spanish Virtual Observatory through project AyA2017-84089, the Galician Regional Government, Xunta de Galicia, through grants ED431B-2021/36, ED481A-2019/155, and ED481A-2021/296, the Centro de Investigación en Tecnologías de la Información y las Comunicaciones (CITIC), funded by the Xunta de Galicia and the European Union (European Regional Development Fund – Galicia 2014-2020 Programme), through grant ED431G-2019/01, the Red Española de Supercomputación (RES) computer resources at MareNostrum, the Barcelona Supercomputing Centre - Centro Nacional de Supercomputación (BSC-CNS) through activities AECT-2017-2-0002, AECT-2017-3-0006, AECT-2018-1-0017, AECT-2018-2-0013, AECT-2018-3-0011, AECT-2019-1-0010, AECT-2019-2-0014, AECT-2019-3-0003, AECT-2020-1-0004, and DATA-2020-1-0010, the Departament d'Innovació, Universitats i Empresa de la Generalitat de Catalunya through grant 2014-SGR-1051 for project 'Models de Programació i Entorns d'Execució Parallels' (MPEXPAR), and Ramon y Cajal Fellowship RYC2018-025968-I funded by MICIN/AEI/10.13039/501100011033 and the European Science Foundation ('Investing in your future'); the Swedish National Space Agency (SNSA/Rymdstyrelsen); the Swiss State Secretariat for Education, Research, and Innovation through the Swiss Activités Nationales Complémentaires and the Swiss National Science Foundation through an Eccellenza Professorial Fellowship (award PCEFP2_194638 for R. Anderson); the United Kingdom Particle Physics and Astronomy Research Council (PPARC), the United Kingdom Science and Technology Facilities Council (STFC), and the United Kingdom Space Agency (UKSA) through the following grants to the University of Bristol, the

**Table 6.** Variability type completeness and contamination estimates of the SOS tables with respect to available cross-matched reference catalogues. For the classification variability types without specific studies, we refer to Rimoldini et al. (2022a). The surveys used as references are SDSS (Lyke et al. 2020), OGLE (Udalski et al. 1992), ASAS-SN (Kochanek et al. 2017), ZTF (Graham et al. 2019).

| Group | Variability type | Catalogue (and region) | Completeness | Contamination |
|---|---|---|---|---|
| AGN | agn | Gaia-CRF3 | 51% | $\leq 5\%$ |
| AGN | agn | SDSS-DR16Q[a] | 47% | $\leq 5\%$ |
| Cepheids | Classical Cepheids | OGLE IV (MW) | $> 86\%$ | $<2\%$ |
| Cepheids | All Cepheids | OGLE IV (LMC & SMC) | $\sim 90\%$ | $<1\%$ |
| Eclipsing binaries | eclipsing_binary | OGLE-IV (LMC/SMC/Bulge) | 33/45/19% | $\sim 5\%$ |
| LPV | long_period_variable | ASAS-SN and OGLE III-LPV [b] | 79–83% | 0.7–2% |
| Microlensing | microlensing | OGLE-IV (Bulge, Disk) | 30-80% | $< 1\%$ |
| Rotation modulation | rotation_modulation | ZTF | 0.4 % [c] | 6% |
| Rotation modulation | rotation_modulation | ASAS-SN | 0.4% | 14% |
| RR Lyrae stars | rrlyr | OGLE-IV (LMC) | 83% | <1.8% |
| RR Lyrae stars | rrlyr | OGLE-IV (SMC) | 94% | <8% |
| RR Lyrae stars | rrlyr | OGLE-IV (Bulge-up) | 79% | <0.15% |
| RR Lyrae stars | rrlyr | OGLE-IV (Bulge-down) | 82% | - |

**Notes.** [a] $+10\,\text{deg} < \text{dec} < +50\,\text{deg}$ and $130\,\text{deg} < \text{ra} < 220\,\text{deg}$. [b] All OGLE-III regions are considered, but filtered by amplitude and specific sky positions. Contamination rate derived from ASAS-SN only. [c] The completeness is strongly dependent on the $G$ magnitude and on the ecliptic latitude (see the discussion in Distefano et al. 2022 for further details).

University of Cambridge, the University of Edinburgh, the University of Leicester, the Mullard Space Sciences Laboratory of University College London, and the United Kingdom Rutherford Appleton Laboratory (RAL): PP/D006511/1, PP/D006546/1, PP/D006570/1, ST/I000852/1, ST/J005045/1, ST/K00056X/1, ST/K000209/1, ST/K000756/1, ST/L006561/1, ST/N000595/1, ST/N000641/1, ST/N000978/1, ST/N001117/1, ST/S000929/1, ST/S000976/1, ST/S000984/1, ST/S001123/1, ST/S001948/1, ST/S001980/1, ST/S002103/1, ST/V000969/1, ST/W002469/1, ST/W002493/1, ST/W002671/1, ST/W002809/1, and EP/V520342/1.

The GBOT programme uses observations collected at (i) the European Organisation for Astronomical Research in the Southern Hemisphere (ESO) with the VLT Survey Telescope (VST), under ESO programmes 092.B-0165, 093.B-0236, 094.B-0181, 095.B-0046, 096.B-0162, 097.B-0304, 098.B-0030, 099.B-0034, 0100.B-0131, 0101.B-0156, 0102.B-0174, and 0103.B-0165; and (ii) the Liverpool Telescope, which is operated on the island of La Palma by Liverpool John Moores University in the Spanish Observatorio del Roque de los Muchachos of the Instituto de Astrofísica de Canarias with financial support from the United Kingdom Science and Technology Facilities Council, and (iii) telescopes of the Las Cumbres Observatory Global Telescope Network.

[1] Institute of Astronomy, University of Cambridge, Madingley Road, Cambridge CB3 0HA, United Kingdom
[2] Université Côte d'Azur, Observatoire de la Côte d'Azur, CNRS, Laboratoire Lagrange, Bd de l'Observatoire, CS 34229, 06304 Nice Cedex 4, France
[3] Department of Astronomy, University of Geneva, Chemin Pegasi 51, 1290 Versoix, Switzerland
[4] Sednai Sàrl, Geneva, Switzerland
[5] Department of Astronomy, University of Geneva, Chemin d'Ecogia 16, 1290 Versoix, Switzerland
[6] INAF - Osservatorio di Astrofisica e Scienza dello Spazio di Bologna, via Piero Gobetti 93/3, 40129 Bologna, Italy
[7] European Space Agency (ESA), European Space Astronomy Centre (ESAC), Camino bajo del Castillo, s/n, Urbanizacion Villafranca del Castillo, Villanueva de la Cañada, 28692 Madrid, Spain
[8] INAF - Osservatorio Astrofisico di Catania, via S. Sofia 78, 95123 Catania, Italy

[9] Dipartimento di Fisica e Astronomia ""Ettore Majorana"", Università di Catania, Via S. Sofia 64, 95123 Catania, Italy

[10] Dpto. de Inteligencia Artificial, UNED, c/ Juan del Rosal 16, 28040 Madrid, Spain

[11] School of Physics and Astronomy, Tel Aviv University, Tel Aviv 6997801, Israel

[12] Department of Particle Physics and Astrophysics, Weizmann Institute of Science, Rehovot 7610001, Israel

[13] RHEA for European Space Agency (ESA), Camino bajo del Castillo, s/n, Urbanizacion Villafranca del Castillo, Villanueva de la Cañada, 28692 Madrid, Spain

[14] Ruđer Bošković Institute, Bijenička cesta 54, 10000 Zagreb, Croatia

[15] INAF - Osservatorio Astrofisico di Torino, via Osservatorio 20, 10025 Pino Torinese (TO), Italy

[16] Konkoly Observatory, Research Centre for Astronomy and Earth Sciences, Eötvös Loránd Research Network (ELKH), MTA Centre of Excellence, Konkoly Thege Miklós út 15-17, 1121 Budapest, Hungary

[17] ELTE Eötvös Loránd University, Institute of Physics, 1117, Pázmány Péter sétány 1A, Budapest, Hungary

[18] Instituut voor Sterrenkunde, KU Leuven, Celestijnenlaan 200D, 3001 Leuven, Belgium

[19] Department of Astrophysics/IMAPP, Radboud University, P.O.Box 9010, 6500 GL Nijmegen, The Netherlands

[20] Max Planck Institute for Astronomy, Königstuhl 17, 69117 Heidelberg, Germany

[21] Institute of Physics, Laboratory of Astrophysics, Ecole Polytechnique Fédérale de Lausanne (EPFL), Observatoire de Sauverny, 1290 Versoix, Switzerland

[22] Porter School of the Environment and Earth Sciences, Tel Aviv University, Tel Aviv 6997801, Israel

[23] Université de Caen Normandie, Côte de Nacre Boulevard Maréchal Juin, 14032 Caen, France

[24] Astronomical Observatory, University of Warsaw, Al. Ujazdowskie 4, 00-478 Warszawa, Poland

[25] University of Vienna, Department of Astrophysics, Türkenschanzstraße 17, A1180 Vienna, Austria

[26] INAF - Osservatorio Astronomico di Capodimonte, Via Moiariello 16, 80131, Napoli, Italy

[27] MTA CSFK Lendület Near-Field Cosmology Research Group, Konkoly Observatory, MTA Research Centre for Astronomy and Earth Sciences, Konkoly Thege Miklós út 15-17, 1121 Budapest, Hungary

[28] Villanova University, Department of Astrophysics and Planetary Science, 800 E Lancaster Avenue, Villanova PA 19085, USA

[29] Institute of Global Health, University of Geneva

[30] Department of Physics and Astronomy G. Galilei, University of Padova, Vicolo dell'Osservatorio 3, 35122, Padova, Italy

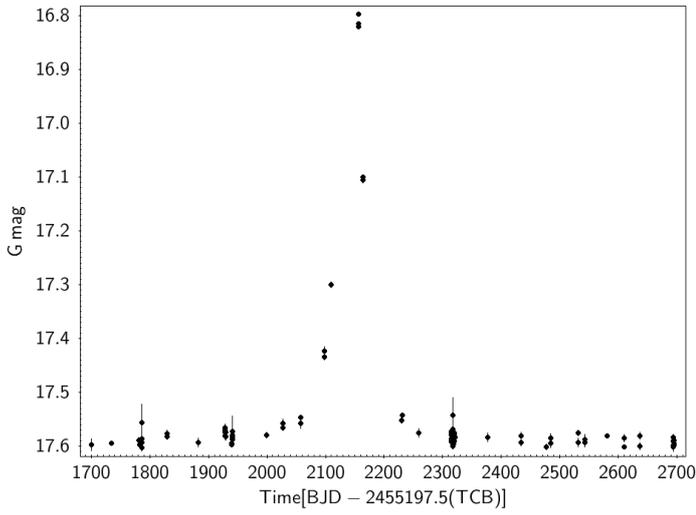
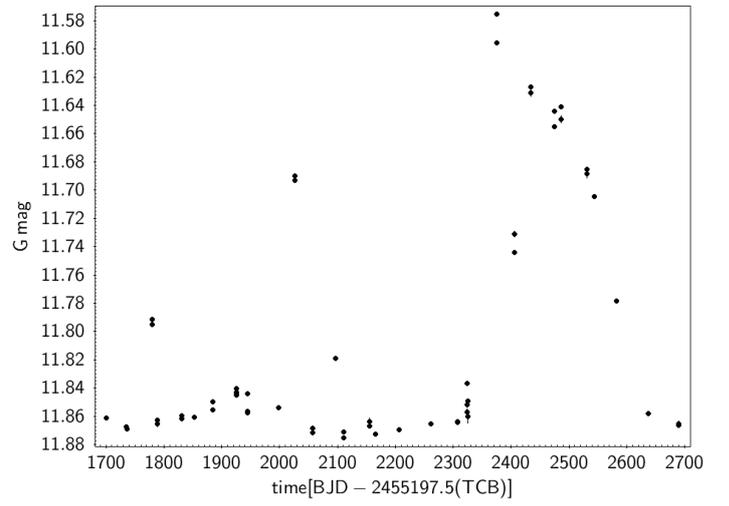
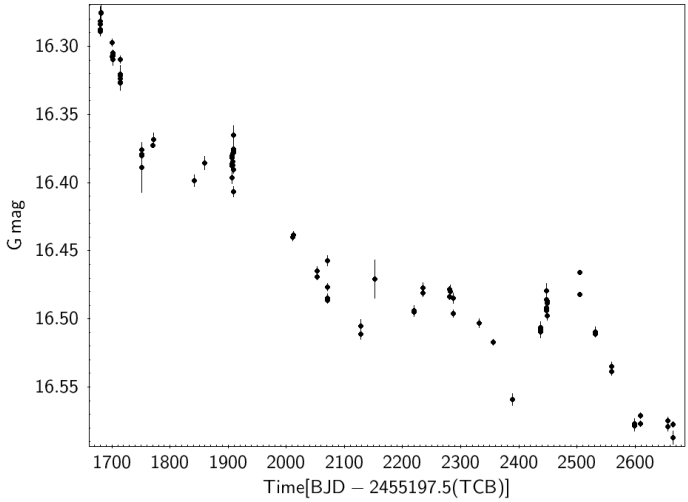
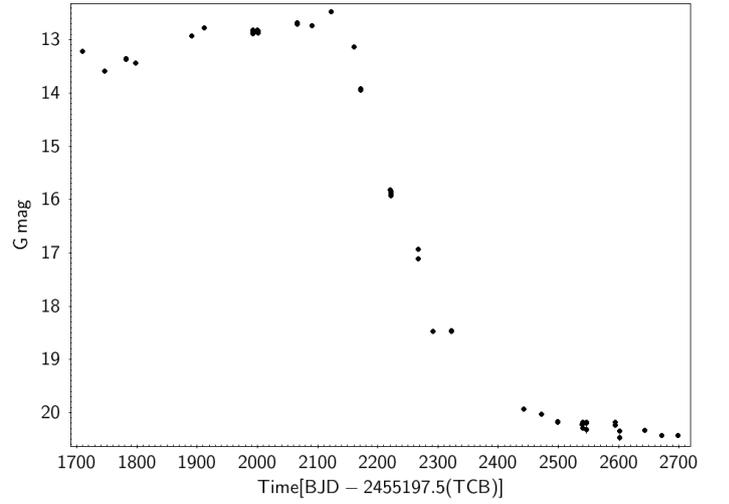
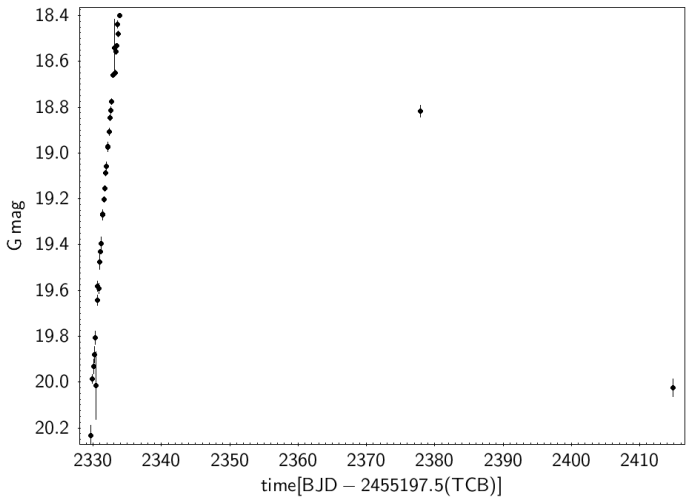
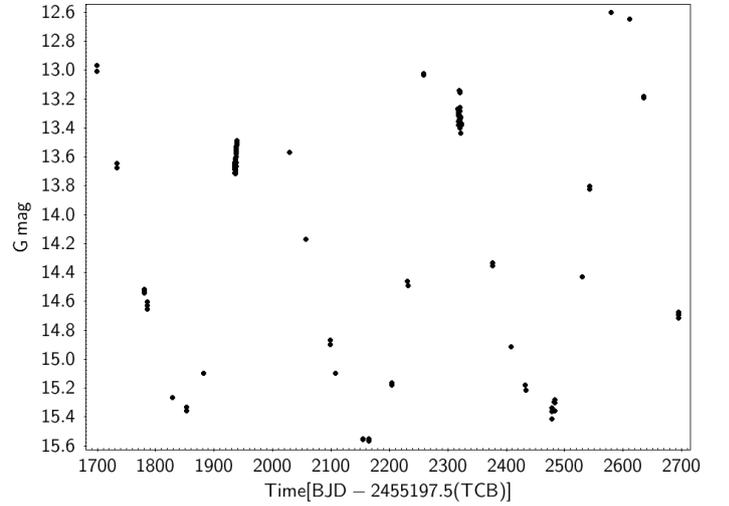

**Fig. A.1.** Light curve of the microlensing event Gaia DR3 3606211681 9613165952, the AGN Gaia DR3 5008130024044117504, the SN Gaia DR3 2859217423244109952.

**Fig. A.2.** Light curve of the Be star Gaia DR3 5862198186620680064, R CBr stars Gaia DR3 6430756164669198336, the long-period variable Gaia DR3 6061778685384497920.

## Appendix A: Atlas of light curves

We present a small atlas of light curves having a long term variability.

## Appendix B: Examples of ADQL queries for the Gaia archive

We give in this section some examples of the queries that were used to produce numbers, tables, figures of this article.

For the Table 1.
All variables (including spurious variability from galaxies) and GAPS:

```
SELECT COUNT(*)
FROM (SELECT source_id, phot_variable_flag,
             in_andromeda_survey
      FROM gaiadr3.gaia_source_lite
      WHERE phot_variable_flag = 'VARIABLE'
          OR in_andromeda_survey = 't'
          OR in_galaxy_candidates = 't') AS s
FULL OUTER JOIN gaiadr3.galaxy_candidates g
    ON g.source_id = s.source_id
WHERE s.phot_variable_flag = 'VARIABLE'
      OR s.in_andromeda_survey = 't'
      OR g.vari_best_class_name = 'GALAXY'
```

All variables: all 'true' variables and galaxies:

```
SELECT COUNT(*)
FROM (SELECT source_id, phot_variable_flag
      FROM gaiadr3.gaia_source_lite
      WHERE phot_variable_flag = 'VARIABLE'
          OR in_galaxy_candidates = 't') AS s
FULL OUTER JOIN gaiadr3.galaxy_candidates g
    ON g.source_id = s.source_id
WHERE s.phot_variable_flag = 'VARIABLE'
      OR g.vari_best_class_name = 'GALAXY'
```

All classifications: classification of 'true' variables and galaxies

```
SELECT COUNT(*)
FROM gaiadr3.vari_classifier_result c
FULL OUTER JOIN gaiadr3.galaxy_candidates g
    ON c.source_id = g.source_id
WHERE g.vari_best_class_name = 'GALAXY'
      OR c.best_class_name IS NOT NULL
```

Vari_summary: all 'true' variables + GAPS (with overlaps), i.e. number of sources having time series:

```
SELECT COUNT(*)
FROM gaiadr3.gaia_source_lite
WHERE has_epoch_photometry = 't'
```

All variable sources:

```
SELECT COUNT(*)
FROM gaiadr3.gaia_source_lite
WHERE phot_variable_flag = 'VARIABLE'
```

Classifications of 'true' variables from the supervised classifier:

```
SELECT COUNT(*)
FROM gaiadr3.vari_classifier_result
```

Classifications of galaxies (artificial variability, in the galaxy candidates table):

```
SELECT COUNT(*)
FROM gaiadr3.galaxy_candidates
WHERE vari_best_class_name = 'GALAXY'
```

GAPS, among which 12 618 published variable sources and 7579 galaxies:

```
SELECT COUNT(*)
FROM gaiadr3.gaia_source_lite
WHERE in_andromeda_survey = 't'
```

Number of SOS sources with period(s) published:

```
<TODO>
```

Variable stars (RR Lyrae and Cepheids) with radial velocity time series:

```
SELECT COUNT(*)
FROM gaiadr3.gaia_source_lite
WHERE has_epoch_rv = 't'
```

For the Table 2.
Union of tables:

```
SELECT source_id FROM gaiadr3.vari_rrlyrae
UNION
SELECT source_id FROM gaiadr3.vari_classifier_result
WHERE best_class_name = 'RR'
```

Intersection of tables:

```
SELECT COUNT(*) FROM gaiadr3.vari_classifier_result AS c
INNER JOIN gaiadr3.vari_agn AS sos
    ON (c.source_id = sos.source_id)
WHERE c.best_class_name = 'AGN'
```

A table where the intersection with another table is removed:

```
SELECT COUNT(*) FROM gaiadr3.vari_classifier_result AS c
LEFT JOIN gaiadr3.vari_agn AS sos
     ON c.source_id = sos.source_id
WHERE c.best_class_name = 'AGN'
      AND sos.source_id IS NULL
```

For the Table 4.
The table extracts columns from two different tables:

```
SELECT s.source_id, sos.frequency, sos.amplitude,
vs.median_mag_bp, vs.median_mag_rp, vs.range_mag_g_fov,
vs.trimmed_range_mag_g_fov, vs.range_mag_bp,
vs.trimmed_range_mag_bp, vs.range_mag_rp,
vs.trimmed_range_mag_rp, vs.std_dev_mag_g_fov,
vs.std_dev_mag_bp, vs.std_dev_mag_rp,
vs.num_selected_g_fov, vs.num_selected_bp,
vs.num_selected_rp
FROM gaiadr3.vari_summary as vs
INNER JOIN gaiadr3.vari_long_period_variable AS sos
    ON (vs.source_id=sos.source_id)
WHERE vs.num_selected_bp > 15
      AND vs.num_selected_rp > 15
```